\documentclass[letterpaper,11pt]{article}
\pdfoutput=1

\usepackage{jheppub}
\usepackage{epsfig}  
\usepackage{graphicx}
\usepackage{hyperref}
\usepackage{color}
\usepackage{float}
\usepackage{amsfonts}
\usepackage{amsmath}
\usepackage{slashed}

\large

\newcommand{\eq}[1]{Eq.~\eqref{eq:#1}}

\newcommand{\fig}[1]{Fig.~\ref{fig:#1}}

\newcommand{\ord}[1]{{\mathcal O}(#1)}

\newcommand{\ORD}[1]{{\mathcal O}\biggl(#1\biggr)}

\newcommand{\df}{\mathrm{d}}
\newcommand{\img}{\mathrm{i}}

\newcommand{\Tau}{\mathcal{T}}

\newcommand{\GeV}{\,\mathrm{GeV}}

\newcommand{\nn}{\nonumber}

\newcommand{\cI}{{\mathcal I}}

\newcommand{\Ecm}{E_\mathrm{cm}}

\newcommand{\cusp}{\mathrm{cusp}}
\newcommand{\cut}{\mathrm{cut}}

\newcommand{\FO}{\mathrm{FO}}
\newcommand{\cm}{\mathrm{cm}}
\newcommand{\nons}{\mathrm{nons}}
\newcommand{\resum}{\mathrm{resum}}

\newcommand{\run}{\mathrm{run}}

\newcommand{\vary}{\mathrm{vary}}

\newcommand{\zero}{{(0)}}
\newcommand{\one}{{(1)}}
\newcommand{\two}{{(2)}}

\begin{document}

\title{Higgs Production at NNLL$'$+NNLO using Rapidity Dependent Jet Vetoes}

\author[a]{Shireen Gangal,}

\author[b]{Jonathan R.~Gaunt,}

\author[c]{Frank J.~Tackmann,}

\author[b]{and Eleni Vryonidou}

\emailAdd{shireen.gangal@theory.tifr.res.in}
\emailAdd{jonathan.richard.gaunt@cern.ch}
\emailAdd{frank.tackmann@desy.de}
\emailAdd{eleni.vryonidou@cern.ch}

\affiliation[a]{Tata Institute of Fundamental Research, Mumbai 400005, India}
\affiliation[b]{CERN Theory Division, 1211 Geneva 23, Switzerland}
\affiliation[c]{Theory Group, Deutsches Elektronen-Synchrotron (DESY), D-22607 Hamburg, Germany\vspace{0.5ex}}

\abstract{
The rapidity-dependent jet veto observables $\Tau_{Bj}$ and $\Tau_{Cj}$ provide a tight jet veto at central rapidity, gradually transitioning to a loose veto at forward rapidities. They divide the phase space into exclusive jet bins in a different way to the traditional jet veto observable $p_{Tj}$, and are advantageous to use under harsh pile-up conditions. We obtain predictions for the $0$-jet gluon-fusion (ggF) Higgs cross section using both of these veto observables at NNLL$'+$NNLO, and compare these predictions to the prior state-of-the-art of NLL$'+$NLO. A significant reduction in perturbative uncertainty is observed going from NLL$'+$NLO to NNLL$'+$NNLO, with the NNLL$'+$NNLO predictions lying inside the uncertainty band of the NLL$'+$NLO predictions. We also investigate the relative sensitivities of ggF Higgs cross sections with $\Tau_{Bj}$, $\Tau_{Cj}$ and $p_{Tj}$ jet vetoes to underlying event and hadronisation effects using an NLO+parton shower calculation. We find that the cross sections with $\Tau_{Bj}$ and $\Tau_{Cj}$ vetoes have a reduced sensitivity to underlying event and hadronisation effects compared to that with a $p_{Tj}$ veto.
}

 \preprint{\begin{flushright}
 CERN-TH-2020-036 \\
 DESY 20-035  \\
 TIFR/TH/20-7 
 \end{flushright}
 }
 
\maketitle 

\section{Introduction}

Jet vetoes find frequent application at the LHC, to separate different types of hard processes and cut away backgrounds. One key process where jet vetoes find application is in Higgs production -- for example, in the $H\to WW^*$ analysis, it is standard to apply a jet veto to reduce the dominant $t\bar{t}$ decay background. 

For maximal background rejection, it is typically advantageous to set the jet veto scale $\Tau^\cut$ to be much smaller than the hard scale of the process $Q$. This induces large logarithms of $\Tau^\cut/Q$ in the perturbative series for the $0$-jet cross section, which can be summed up to all orders to obtain precise predictions \cite{Stewart:2009yx,Berger:2010xi}.

The standard jet variable by which jets are currently classified and vetoed is the transverse momentum $p_{Tj}$ of a jet. One identifies all jets with a given radius $R$ in an event, computes $p_{Tj}$ for each jet, and vetoes the event if the $p_{Tj}$ for any jet exceeds the veto scale $p_{Tj}^\cut$. Predictions for the $0$-jet Higgs production cross section with a jet veto imposed via $p_{Tj}$ have been obtained with a resummation of $p_{Tj}/m_H$ logarithms up to the NNLL$'$ order, matched to either NNLO or N$^3$LO fixed order perturbation theory \cite{Banfi:2012yh, Banfi:2012jm, Becher:2012qa, Becher:2013xia, Stewart:2013faa, Banfi:2015pju}. However, there are some drawbacks to using this variable. In harsh pile-up conditions, as are nowadays encountered at the LHC, it is hard to identify (and veto) jets with small $p_{Tj}$ at large rapidity. This is due to the lack of tracking information at large $|\eta_j| \ge 2.5$, meaning that at large rapidities it is difficult to disentangle small $p_{Tj}$ jets from the primary process from pile-up jets. One option is simply to raise the overall cut on $p_{Tj}$, but then the discriminating power of the jet veto is reduced. An alternative possibility is to consider a step-like jet veto, with a tight veto below some value of $|\eta|$, and a weaker veto above this value. The theoretical description of such step-like vetoes is discussed in Ref.~\cite{Michel:2018hui}.

A final possibility, proposed in Refs.~\cite{Tackmann:2012bt, Gangal:2014qda} and which we focus on here, is to have a jet veto that is a smooth function of rapidity, being tight at central rapidities and gradually weakening as one goes forward. Four jet-veto variables of this kind were considered in Ref.~\cite{Gangal:2014qda}:
\begin{align} \label{eq:TauBCdefs}
&\mathcal{T}_{Bj} = m_{Tj}e^{-|y_j - Y|} 
&&\mathcal{T}_{Cj} = \dfrac{m_{Tj}}{2\,\text{cosh}\left(y_j-Y\right)}
\\ \label{eq:TauBCcmdefs}
&\mathcal{T}_{Bj\cm} = m_{Tj}e^{-|y_j|} 
 &&\mathcal{T}_{Cj\cm} = \dfrac{m_{Tj}}{2\,\text{cosh}\left(y_j\right)}
\end{align}
where $y_j$ the rapidity of the jet and $Y$ the rapidity of the Higgs. The transverse mass of the jet $m_{Tj} = \sqrt{m_j^2 + p_{Tj}^2}$ and is close to $p_{Tj}$ for small jet radius $R$. The quantity $\mathcal{T}_{Bj(\cm)}$ has the same rapidity weighting as the global beam thrust hadronic event shape \cite{Stewart:2009yx, Stewart:2010pd}, whilst $\mathcal{T}_{Cj(\cm)}$ has the same rapidity weighting as the $C$-parameter defined for $e^+e^- \to \mathrm{hadrons}$. Resummation of veto logarithms for all of these observables can be achieved using the same framework (see Ref.~\cite{Gangal:2014qda} for details).

The veto observables in Eqs.~\eqref{eq:TauBCdefs} and \eqref{eq:TauBCcmdefs} are closely linked, differing only in their reference point from which to define whether jets are `central' and `forward'; the observables in Eq.~\eqref{eq:TauBCdefs} do it with respect to the Higgs rapidity, whilst those in Eq.~\eqref{eq:TauBCcmdefs} do it with respect to the centre of momentum of the $pp$ collision. In terms of avoiding the issues with forward pile-up jets, the latter alternative will be superior, although there should not be a drastic difference in performance between the two types of vetoes at the LHC, since the majority of Higgs bosons are produced at rather low rapidities. Here, we choose to focus on the observables in Eq.~\eqref{eq:TauBCdefs}, motivated in part by the fact that these seem to be the preferred choice from an experimental perspective: in Ref.~\cite{Aad:2014lwa} the $H+0$-jet cross section differential in $\Tau_{Cj}$ was measured in the ATLAS $H \to \gamma \gamma$ analysis.

Apart from the above considerations, given the general utility of jet binning it is clearly beneficial to have more than one way of dividing phase space up into jet bins -- for certain analyses it may be advantageous to use $\mathcal{T}_{B/Cj}$ rather than $p_{Tj}$. Vetoed cross sections using $\mathcal{T}_{B/Cj}$ probe QCD radiation in a quite different way from cross sections with a $p_{Tj}$ veto\footnote{Technically: $p_{Tj}$ veto cross sections are SCET$_{\mathrm{II}}$ observables whilst $\mathcal{T}_{B/Cj}$ veto cross sections are SCET$_{\mathrm{I}}$ observables.}, so studying and measuring these observables is also of interest in terms of testing our understanding of QCD radiation.  

The factorisation framework to resum veto logarithms for colour-singlet $0$-jet processes with a jet veto imposed via $\Tau_{B/Cj}$ was established in Ref.~\cite{Tackmann:2012bt, Gangal:2014qda} within soft collinear effective theory (SCET). For the gluon-fusion (ggF) Higgs cross section, predictions for the $0$-jet cross section with a $\Tau_{B/Cj}$ jet veto have been obtained at NLL$'$+NLO \cite{ Gangal:2014qda}. In Ref.~\cite{Gangal:2016kuo}, the resummation ingredients (two-loop beam and soft functions) required for the NNLL$'$ resummation of colour-singlet $0$-jet processes with a $\Tau_{B/Cj}$ jet veto were computed.

The main goal of the present paper is to elevate the precision of the $0$-jet ggF Higgs production predictions, for both $\Tau_{Bj}$ and $\Tau_{Cj}$ to NNLL$'$ in resummed logarithms of $\Tau_{B/Cj}/m_H$, matched to NNLO fixed-order perturbation theory. We will include finite bottom and top quark mass effects up to two loops (NLO) in the Higgs production process, and resum time-like logarithms
in the $gg\to H$ form factor to all orders \cite{Parisi:1979xd, Sterman:1986aj, Magnea:1990zb, Ahrens:2008qu, Ahrens:2008nc, Ebert:2017uel}. Note that here we do not perform a resummation of logarithms of $R$ (studied in Refs.~\cite{Dasgupta:2014yra, Banfi:2015pju, Kang:2016mcy, Dai:2016hzf}), nor logarithms of $m_b/m_H$ (studied in Refs.~\cite{Melnikov:2016emg, Liu:2018czl, Liu:2019oav}). 

A further goal is to investigate the relative sensitivity of the $0$-jet ggH cross sections with $\Tau_{Bj}$, $\Tau_{Cj}$ and $p_{Tj}$ vetoes to the effects of hadronisation and underlying event (UE). These effects are difficult to describe from first principles and are typically modelled, so a minimal sensitivity to them is preferable. This investigation will be performed using a NLO + parton shower (PS) set-up, specifically {\sc Madgraph\_aMC@NLO} \cite{Alwall:2014hca} + Pythia8 \cite{Sjostrand:2014zea}. We find that the cross sections with $\Tau_{Bj}$ and $\Tau_{Cj}$ vetoes are rather less sensitive to both UE and hadronisation than that with a $p_{Tj}$ veto, which constitutes another advantage of using these veto observables.

This paper is organised as follows. In section \ref{sec:I} we review the SCET factorization formula derived in Refs.~\cite{Tackmann:2012bt, Gangal:2014qda} for the $H+0$-jet $\Tau_{B/Cj}$ cross section, and outline the steps and necessary ingredients needed to obtain this cross section at NNLL$'+$NNLO. In section \ref{sec:II} we give details of the procedure we use to determine the perturbative uncertainty in our results. Section \ref{sec:III} contains our results for the $H+0$-jet $\Tau_{B/Cj}$ cross section at NNLL$'+$NNLO, and we investigate the effects of UE and hadronisation on the $H+0$-jet $\Tau_{B/Cj}$ and $p_{Tj}$ cross sections using an NLO+PS set-up in section \ref{sec:IV}. Finally we conclude in section \ref{sec:V}. 

\section{Factorization for the $H+0$-jet cross section }
\label{sec:I}

The full $pp \to H + 0$-jet cross section with a cut on the rapidity-dependent observable $\Tau_{fj} < \Tau^\cut$ ($f=B,C$) is given by,  
\begin{align} \label{eq:fullXsec}
\sigma_0 (\Tau_{fj} \!< \Tau^\cut)
&= \sigma_0^\resum (\Tau_{fj} \!< \Tau^\cut)
+ \sigma_0^\nons (\Tau_{fj} \!< \Tau^\cut)
\,,\end{align}
where the first term contains the resummed logarithms of $\Tau^\cut/m_H$, and dominates for small values of $\Tau^\cut$, while the second term contains the nonsingular corrections which are
suppressed by $O(\Tau^\cut/m_H)$ and become important at large $\Tau^\cut$.  
The resummed $H$+0-jet cross section for $\Tau_{fj} < \Tau^\cut$ can be factorized as follows, \cite{Tackmann:2012bt, Gangal:2014qda}
\begin{align} \label{eq:TauBCfacto}
\sigma_0^\resum (\Tau_{fj} \!< \Tau^\cut)
= & \, \sigma_B H_{gg}(m_t, m_b, m_H^2, \mu)\, B_g(m_H \Tau^{\cut},x_a,R,\mu)
B_g(m_H \Tau^{\cut},x_b,R,\mu)\, 
\nn\\
& \times S_{f}(\Tau^{\cut},R,\mu) + \sigma_0^{\text{Rsub}}(\Tau_{fj} \!< \Tau^\cut, R)
\,,\end{align}
where
\begin{align}
x_{a,b} = \frac{m_H}{\Ecm}\,e^{\pm Y}
\,,\quad
\sigma_B = \frac{\sqrt{2} G_F\, m_H^2}{576 \pi \Ecm^2}
\,.\end{align}
The hard function denoted by $H_{gg}$ contains the hard virtual corrections and is obtained by matching QCD onto the operator $O_{ggH}$ in SCET. The gluon beam function $B_g$ describes the collinear initial state radiation from the
incoming gluons, while $S_{f}$ encodes the contribution from the soft radiation across the entire event.  The term  $\sigma_0^{\text{Rsub}}$ contains $O(R^2)$ corrections arising due to the
clustering of two independent collinear or soft particles into one jet; following Refs.~\cite{Tackmann:2012bt, Stewart:2013faa} we separate these off and treat them separately\footnote{This scheme is the second one discussed in section 3.2 of Ref.~\cite{Gangal:2016kuo}. This implies that in the two-loop $B$ and $S$, we do not include the $\Delta I_{\text{indep}}, S\hspace{-0.2mm}C$ and  $\Delta S_{f,\text{indep}}$ terms computed in Ref.~\cite{Gangal:2016kuo}, and remove the $C_A^2 R^2$ term associated with independent emission contributions from the anomalous dimensions of $B$ and $S$.}. 

The hard function is the IR-finite part of the $\overline{\text{MS}}$ renormalized $ggH$ form factor, and is expressed in terms of the
matching coefficient $C_{ggH}$ as,
\begin{align}
H_{gg}(m_t, m_b, m_H^2, \mu) = |C_{ggH}(m_t, m_b, m_H^2, \mu)|^2
\,.\end{align}

At NNLL$^\prime$, we require the hard function up to NNLO:
\begin{align} \label{eq:Hfuncexp}
H_{gg}=& |C^{(0)}_{ggH}|^2+ \frac{\alpha_s(\mu)}{4\pi}2\mathrm{Re}\left[C^{(1)}_{ggH}C^{(0)*}_{ggH}\right]
\\ \nonumber
&+\frac{\alpha_s^2(\mu)}{(4\pi)^2}\left(2\mathrm{Re}\left[C^{(2)}_{ggH}C^{(0)*}_{ggH}\right]+|C^{(1)}_{ggH}|^2 \right)+...
\,,\end{align}
where the $C^{(0,1,2)}_{ggH}$ are the perturbative expansion coefficients of $C_{ggH}$:
\begin{align}
C_{ggH} = C^{(0)}_{ggH} + \dfrac{\alpha_s}{4\pi} C^{(1)}_{ggH} + \left(\dfrac{\alpha_s}{4\pi}\right)^2 C^{(2)}_{ggH}+... \,.
\end{align}
Note that since $C_{ggH}$ begins at $\mathcal{O}(\alpha_s)$ we absorb one power of $\alpha_s$ into each coefficient $C^{(0,1,2)}_{ggH}$.

In $C^{(0)}$ and $C^{(1)}$ we include the top and bottom loops with the full mass dependence using the results of Refs.~\cite{Harlander:2005rq, Anastasiou:2006hc}. For the NNLO coefficient $C^{(2)}$, we use the NNLO $m_t\to\infty$ result extracted from Refs.~\cite{Pak:2009bx, Harlander:2009bw}, reweighted by the ratio of $C^{(0)}$ with finite $m_t$ to $C^{(0)}$ with $m_t\to\infty$ (both with $m_b=0$). This approach to the NNLO term is the same as in Ref.~\cite{Berger:2010xi}.

The beam functions $B_g$ are the same for both the $\Tau_{Bj}$ and $\Tau_{Cj}$ observables as they describe collinear emissions in the regions of large forward rapidities where the measurement function for both these observables is the same. They can be computed as a convolution between the perturbative matching kernel $\mathcal{I}_{ij}$ and the standard parton distribution functions (PDFs) as follows \cite{Stewart:2009yx, Stewart:2010qs},
\begin{align}  \label{eq:BOPE}
B_{g}(t^{\cut}\!,x,R,\mu) &= \sum_j \!\int^1_x \!\! \frac{\df z}{z}\, \mathcal{I}_{gj}(t^{\cut}\!,z,R,\mu) f_{j}\Bigl(\frac{x}{z},\mu \Bigr)
\times
\biggl[1 + \ORD{\frac{\Lambda_\mathrm{QCD}^2}{t^{\cut}}} \biggr]
\,.\end{align}
For NNLL$'$ precision we require the matching kernel up to $\ord{\alpha_s^2}$:
\begin{align} \label{eq:Iij}
\cI_{gj}(t,z,\mu)
&= \cI_{gj}^\zero(t^\cut,x,\mu)
   + \frac{\alpha_s(\mu)}{4\pi}\, \cI_{gj}^\one(t^\cut,x,\mu)
   + \frac{\alpha_s^2(\mu)}{(4\pi)^2}\, \cI_{gj}^\two(t^\cut,x,\mu,R)
   + \ord{\alpha_s^3}
\,.\end{align}
The one-loop coefficients $\cI^\one$ are given in Ref.~\cite{Gangal:2014qda} and are equal to the cumulants of the one-loop virtuality-dependent beam function matching coefficients computed in Ref.~\cite{Berger:2010xi}. One can write the two-loop coefficients $\cI^\two$ as follows:
\begin{align}
\cI_{gj}^\two(t^\cut,x,\mu,R) = \cI_{G,gj}^\two(t^\cut, x,\mu) + \Delta \cI_{gj}^\two (t^\cut,x,\mu,R)
\end{align}
where the $\cI_{G}^\two$ are the cumulants of the two-loop virtuality-dependent beam function matching coefficients computed in Ref.~\cite{Gaunt:2014cfa}, and the expressions for $\Delta \cI^\two$ are given in section 3.1.2 of Ref.~\cite{Gangal:2016kuo}.

The soft function is defined as a vacuum matrix element of the product of soft Wilson lines along the two incoming gluon directions, with a measurement function imposing $\Tau_{fj} < \Tau^\cut$. As with the other ingredients, we require its expansion to NNLO:
\begin{align}
S_f(\Tau^\cut, R,\mu) = 1 +  \frac{\alpha_s(\mu)}{4\pi} S_f^\one (\Tau^\cut, \mu) +   \frac{\alpha_s^2(\mu)}{(4\pi)^2} S_f^\two (\Tau^\cut, R,\mu) + \ord{\alpha_s^3}
\end{align}
The one-loop soft functions $S_f^\one$ for $\Tau_{Bj}$ and $\Tau_{Cj}$ are equal to the cumulants of the one-loop soft functions for thrust and $C$-parameter respectively; the expressions for both observables are given in Ref.~\cite{Gangal:2014qda}. Let us decompose the two-loop soft function as follows:
\begin{align} \label{eq:softsplit}
 S_f^\two (\Tau^\cut, R,\mu)  = S_{G,f}^{\two,\text{non-Ab}} (\Tau^\cut, R,\mu) + \frac{1}{2}\left[ S_f^\one (\Tau^\cut, \mu) \right]^2 +  \Delta S_f^\two  (\Tau^\cut, R,\mu) 
\end{align}
The quantity $S_{G,f}^{\two,\text{non-Ab}}$ is the cumulant of the `non-Abelian' part of the  two-loop thrust/$C$-parameter soft function. Expressions for $\Delta S_f^\two$ are given in section 3.1.1 of Ref.~\cite{Gangal:2016kuo}. Analytic results for $S_{G,B}^{\two,\text{non-Ab}}$ can be extracted from the results in Refs.~\cite{Kelley:2011ng, Monni:2011gb, Hornig:2011iu, Gaunt:2015pea}, whilst for $S_{G,C}^{\two,\text{non-Ab}}$ we extracted numerical results for $n_f=5$ from Ref.~\cite{Hoang:2014wka}. In both cases the results are of the form:
\begin{align}
S_{G,f}^{\two,\text{non-Ab}} =&  a_{G,f}C_A + b_{G,f}C_A \ln\left(\dfrac{\Tau^\cut}{\mu}\right) - \frac{8}{9}C_A\left( 67C_A - 3C_A\pi^2 - 20 n_f T_F\right) \ln\left(\dfrac{\Tau^\cut}{\mu}\right)^2 
\nonumber \\ 
&+ \frac{16}{9}C_A \left( 11C_A - 4n_fT_F\right) \ln\left(\dfrac{\Tau^\cut}{\mu}\right)^3
\end{align}
where $a_{G,f}$ and $b_{G,f}$ are given by:
\begin{align}
&a_{G,C}|_{n_f=5} = 124.075  \\
&b_{G,C}|_{n_f=5} = - 265.650 \\
&a_{G,B} = \dfrac{1}{810}\left[ 20n_fT_F(40+111\pi^2 - 1044\zeta_3) +C_A(-21400-5025\pi^2 + 396\pi^4 + 57420\zeta_3)\right] \\
&b_{G,B} = -\dfrac{4}{27}\left[ n_fT_F (112 -12\pi^2) + C_A( 33\pi^2-404+378\zeta_3)   \right]
\end{align}
Numerical results for the two-loop soft functions $S_{f}^\two(\Tau^\cut, R,\mu) $ have also been obtained using SoftSERVE \cite{Bell:2018jvf, Bell:2018vaa, Bell:2018oqa}, where these results agree with those obtained using the procedure above. 

The hard, beam and soft functions are evaluated at their natural scales $\mu_H \sim m_H$, $\mu_B \sim \sqrt{m_H \Tau^\cut}$ and $\mu_S \sim \Tau^\cut$, to minimize the logarithms they contain, and then RG evolved to a common scale $\mu_{\mathrm{FO}} \sim m_H$ which sums the large logarithms. These satisfy RG equations with a multiplicative form due to the cumulant nature of the jet veto observables \cite{Tackmann:2012bt}:
\begin{align} \label{eq:anomdims}
\mu \frac{\df}{\df\mu} \ln\bigl[C_{ggH} (m_t,m_H^2,\mu)\bigr]
&= \gamma_{H}^g(m_H^2,\mu)
\,, \nn\\
\mu \frac{\df}{\df\mu} \ln\bigl[B_g (t^\cut,x,R,\mu) \bigr]
&= \gamma_{B}^g(t^\cut,R,\mu)
\,, \nn\\
\mu \frac{\df}{\df\mu} \ln\bigl[ S_g^{B,C} (\Tau^\cut\!,R,\mu) \bigr]
&= \gamma_{S}^g(\Tau^\cut,R,\mu)
\,,\end{align}
where the anomalous dimension has a generic form consisting of the $\mu$-dependent cusp part and the non-cusp part as follows,
\begin{align} \label{eq:anomdimsstructure}
\gamma_H^g(m_H^2,\mu) &= \Gamma_\cusp^g[\alpha_s(\mu)] \ln{\frac{-m_H^2\!-\!{\rm i}0}{\mu^2}} + \gamma_H^g[\alpha_s(\mu)]
\,, \nn \\
\gamma_B^g(t^\cut\!,R,\mu) &= -2 \Gamma_\cusp^g[\alpha_s(\mu)] \ln{\frac{t^\cut}{\mu^2}} + \gamma_B^g[\alpha_s(\mu), R]
\,,\nn \\
\gamma_S^g(\Tau^{\cut}\!,R,\mu) &= 4\Gamma_\cusp^g[\alpha_s(\mu)] \ln {\frac{\Tau^{\cut}}{\mu}} + \gamma_S^g[\alpha_s(\mu), R]
\,.\end{align}

The hard matching coefficient contains double logarithms of $(-m_H^2- {\rm i} 0)/\mu_H^2$, so we choose $\mu_H \sim -{\rm i} m_H$ in order to avoid large left
over logarithms $\ln^2({-1 - \rm{i}0}) = -\pi^2$. Choosing an imaginary hard scale ensures that the logarithms are fully resummed and results in better perturbative convergence. For NNLL$'$ resummation we require the cusp anomalous dimension $\Gamma_\cusp$ up to 3 loops, and the non-cusp anomalous dimensions up to 2 loops. The expression for the former is given, for example, in Ref.~\cite{Berger:2010xi} (using the results of Refs.~\cite{Moch:2004pa, Korchemsky:1987wg}). The two-loop expression for $\gamma^g_H[\alpha_s(\mu)]$ can be found in the same paper (using the results of Refs.~\cite{Idilbi:2005ni, Moch:2005tm, Idilbi:2006dg}). The dependence on jet radius enters through the two-loop non-cusp anomalous dimension of the beam and soft function, which can be split into a global and an $R$-dependent part,
\begin{align}
\gamma^g_S [\alpha_s(\mu), R] &= \gamma^g_{G,S} [\alpha_s(\mu)] + \Delta \gamma^g_S [\alpha_s(\mu), R] \nn\\
\gamma^g_B [\alpha_s(\mu), R] &= \gamma^g_{G,B} [\alpha_s(\mu)] + \Delta \gamma^g_B [\alpha_s(\mu), R]
\end{align}
The global non-cusp anomalous dimensions $\gamma^g_{G,S}[\alpha_s(\mu)]$ and $\gamma^g_{G,B}[\alpha_s(\mu)]$  are those of the beam thrust soft and beam functions, and are given up to 2 loops in Ref.~\cite{Berger:2010xi}. The $R$-dependent correction term for the soft function anomalous dimension, $\Delta \gamma^g_S[\alpha_s(\mu), R]$, is given in section 3.1.1 of Ref.~\cite{Gangal:2016kuo}; RGE consistency demands $\Delta \gamma^g_B [\alpha_s(\mu), R] = -\frac{1}{2} \Delta \gamma^g_S [\alpha_s(\mu), R]$.

The solution of the RGE (\eq{anomdims}) has a similar structure for the hard, beam and soft functions. For the beam function, solving the RGE yields,
\begin{align} \label{eq:Bresum}
B_g(t^{\cut}\!,x,R,\mu) = U_B(t^{\cut}\!,\mu_B, \mu)\, B_g(t^{\cut}\!,x,R,\mu_B)
\end{align}
with the evolution factor given by
\begin{align} \label{eq:evolveB}
U_B(t^\cut\!,\mu_B,\mu) &= e^{K_B(\mu_B, \mu)}\Big(\frac{t^\cut}{\mu_B^2}\Big)^{\eta_B(\mu_B, \mu)}
\,.\end{align}
Explicit expressions for the evolution of the hard and soft functions along with the various factors relevant for the resummation at NNLL$^\prime$ can be found in Refs.~\cite{Berger:2010xi, Gangal:2014qda}. \\

The contribution $\sigma_0^\mathrm{Rsub}$ in Eq.~\eqref{eq:TauBCfacto} takes account of the $\mathcal{O}(R^2)$ corrections for the clustering of independent emissions. The form of this term is as follows \cite{Banfi:2012jm, Stewart:2013faa,Gangal:2016kuo}:
\begin{align} \nn
\sigma_0^\mathrm{Rsub} =& \dfrac{\alpha_s^2(\mu_\mathrm{avg})}{(4\pi)^2}H^\zero_{gg}\, U_{\rm total}(\Tau^\cut, \mu_H, \mu_B,\mu_S, \mu_\text{FO}) \, \times
\\
&\Big[ 
\left\{ f_g\,( x_a, \mu_B)\,f_j\,(x_b, \mu_B)\otimes\left(\Delta I^\two_{gj, \mathrm{indep}} (x_b, \mu_\mathrm {avg},R) + S\hspace{-0.5mm}C_{gj}^\two  (x_b, \mu_\mathrm {avg},R)\right)  + (x_a \leftrightarrow x_b) \right\}    \nn\\
& \,\, + f_g\,( x_a, \mu_B)\,f_g\,(x_b, \mu_B) \,\Delta S^\two_{f,\mathrm {indep}}\, (\Tau^\cut , \mu_\mathrm{avg},R ) \Big]
\end{align}
where 
\begin{align}
U_{\rm total}(\Tau^\cut, \mu_H, \mu_B,\mu_S, \mu)  = U_S (\Tau^\cut, \mu_S, \mu) \times U_B^2 (m_H\Tau^\cut, \mu_B, \mu) \times U_H (\mu_H, \mu)
\end{align} 
is the total NNLL$^\prime$ evolution kernel, and $\mu_{\rm avg} = \sqrt{\mu_B \mu_S}$. Since the $\mathcal{O}(\alpha_s^2)$ corrections in the square brackets come from soft or collinear emissions, we choose to evaluate them at the geometric mean of the soft and beam scales, as in Ref.~\cite{Stewart:2013faa}. The expressions for these corrections can be found in section 3.2 of Ref.~\cite{Gangal:2016kuo}.\\

The nonsingular cross section computed at fixed order up to NNLO at scale $\mu_{\text{FO}}$ can be obtained by expanding Eq.~\eqref{eq:fullXsec} up to NNLO with all scales set to $\mu_{\text{FO}}$. In Eq.~\eqref{eq:TauBCfacto} we expand the product of fixed order contributions to the $H, B$ and $S$ factors to $\mathcal{O}(\alpha_s^2$), such that setting all scales to $\mu_{\text{FO}}$ automatically results in the NNLO expansion. Thus:
\begin{align} \label{eq:nonsingeq1}
\sigma_0^{\nons, \text{NNLO}} (\Tau_{fj} \!< \Tau^\cut,\mu_{\text{FO}})
=& 
\sigma^{\mathrm{FO}, \text{NNLO}}_0 (\Tau_{fj} \!< \Tau^\cut) \\ \nn &- \sigma_0^{\resum, \text{NNLL}'} (\Tau_{fj} \!< \Tau^\cut, \mu_B = \mu_S = \mu_H =\mu_{\mathrm{FO}})
\,,\end{align}
We compute the first term on the right hand side at NNLO as follows:
\begin{align}
\sigma^{\mathrm{FO, NNLO}}_{0} (\Tau_{fj} \!< \Tau^\cut) = \sigma^{\mathrm{FO, NNLO}}_{\ge 0}  -\sigma^{\mathrm{FO, NLO}}_{\ge 1} (\Tau_{fj} \!> \Tau^\cut)
\end{align}
where $\sigma^{\mathrm{FO, NNLO}}_{\ge 0}$ is the full NNLO Higgs production cross section, and $\sigma^{\mathrm{FO, NLO}}_{\ge 1} (\Tau_{fj} \!> \Tau^\cut)$ is the NLO $H+j$ cross section with the given cut.

We improve the NNLO nonsingular piece by including the  resummation of time-like logarithms in this piece as well. The expression for the final NNLO nonsingular piece that we include is then \cite{Berger:2010xi, Stewart:2013faa}: 
\begin{align} 
\sigma_0^{\nons, \text{NNLO}+\pi^2} (\Tau_{fj} \!< \Tau^\cut) =& \left[ \sigma_0^{\nons, \text{NNLO}} (\Tau_{fj} \!< \Tau^\cut) - \dfrac{\alpha_s(\mu_{\text{FO}})C_A\pi^2}{2\pi}\sigma_0^{\nons, \text{NLO}} (\Tau_{fj} \!< \Tau^\cut) \right]
\nn \\
& \times U_H(m_H^2,-i\mu_{\text{FO}},\mu_{\text{FO}})
\end{align}

In practice, the computation of $\sigma_0^{\nons, \text{NNLO}}$ is done with finite $m_b, m_t$ up to NLO, and with $m_b \to 0, m_t \to \infty$ in the NNLO coefficient. This means that in the $\sigma_0^{\resum, \text{NNLL}'}$ term of Eq.~\eqref{eq:nonsingeq1} we use a hard function $H_{gg}$ which is given by using Eq.~\eqref{eq:Hfuncexp} but with the $m_t \to \infty$ limit in the NNLO coefficient. The NNLO Higgs cross section is obtained using HNNLO \cite{Catani:2007vq, Grazzini:2008tf, Grazzini:2013mca}, whilst the $H+j$ cross section for  $\Tau_{fj} \!> \Tau^\cut$ is generated using {\sc Madgraph5\_aMC@NLO} \cite{Alwall:2014hca, Hirschi:2015iia}. The computation of  $\sigma_0^{\nons, \text{NLO}}$ is done with finite $m_t,m_b$ using similar techniques. This procedure implies that for $\Tau^\cut \to \infty$, we reproduce the NNLO cross section including the resummation of time-like logarithms \cite{Ebert:2017uel}, which is numerically very close to the N$^3$LO total cross section \cite{Anastasiou:2016cez, Mistlberger:2018etf}.

\section{Profile scales and Perturbative Uncertainties}
\label{sec:II}

We now discuss how we choose the beam and soft scales as a function of $\Tau^\cut$, which, following a standard convention for SCET computations, we refer to as the profile scales  \cite{Ligeti:2008ac, Abbate:2010xh}. At small $\Tau^\cut \ll m_H $, logarithms of $\Tau^\cut/m_H $ are large, Eq.~\eqref{eq:TauBCfacto} applies, and we should set the beam and soft scales to their canonical values $\mu_B \sim \sqrt{m_H \Tau^\cut}, \mu_S \sim  \Tau^\cut$ to resum these logarithms appropriately. This is referred to as the resummation region. At large $\Tau^\cut \gtrsim m_H $, we enter the fixed-order region, in which the veto is sufficiently loose that the fixed-order formula applies, and we should set $\mu_B \sim m_H, \mu_S \sim  m_H$. In the intermediate region, referred to as the transition region, one should ensure a smooth transition between these two behaviours,  which we achieve through a suitable choice of profile scales.

One can determine the locations of the boundaries between these regions by plotting the fixed-order singular and nonsingular contributions to the fixed-order NNLO cross section differential in $\Tau_{fj}$. The singular piece directly follows from taking the derivative of Eq.~\eqref{eq:TauBCfacto} with respect to $\Tau^\cut$, and contains the leading small $\Tau_{fj}$ terms $\sim \ln^n(\Tau_{fj}/m_H)/\Tau_{fj}$ (corresponding to the $\ln^n(\Tau^\cut/m_H)$ terms in the cumulant). The nonsingular contains the remainder of the NNLO cross section, obtained in practice by subtracting the singular contribution from a NLO $H+j$ computation of the $\Tau_{fj}$ spectrum from {\sc MadGraph5\_aMC@NLO}. At small $\Tau_{fj}$ this should diverge at most as $\ln^n(\Tau_{fj}/m_H)$. The resummation region should be defined by where the singular cross section greatly exceeds the nonsingular one, and we should enter the fixed-order region when the singular and nonsingular cross sections become of the same size. 

\begin{figure}[t]
\includegraphics[width=0.46\textwidth]{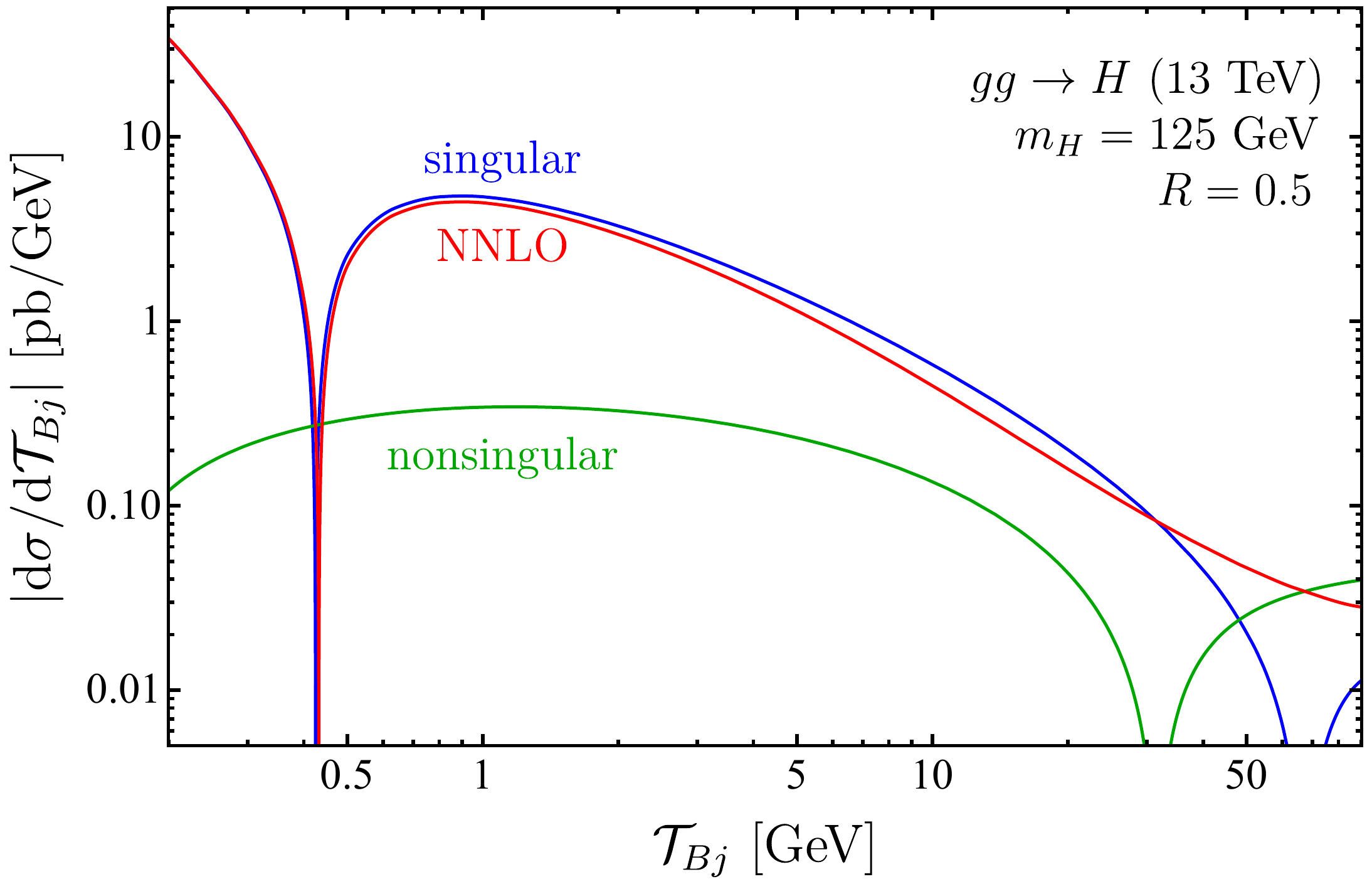}
\hspace{0.02\textwidth}
\includegraphics[width=0.46\textwidth]{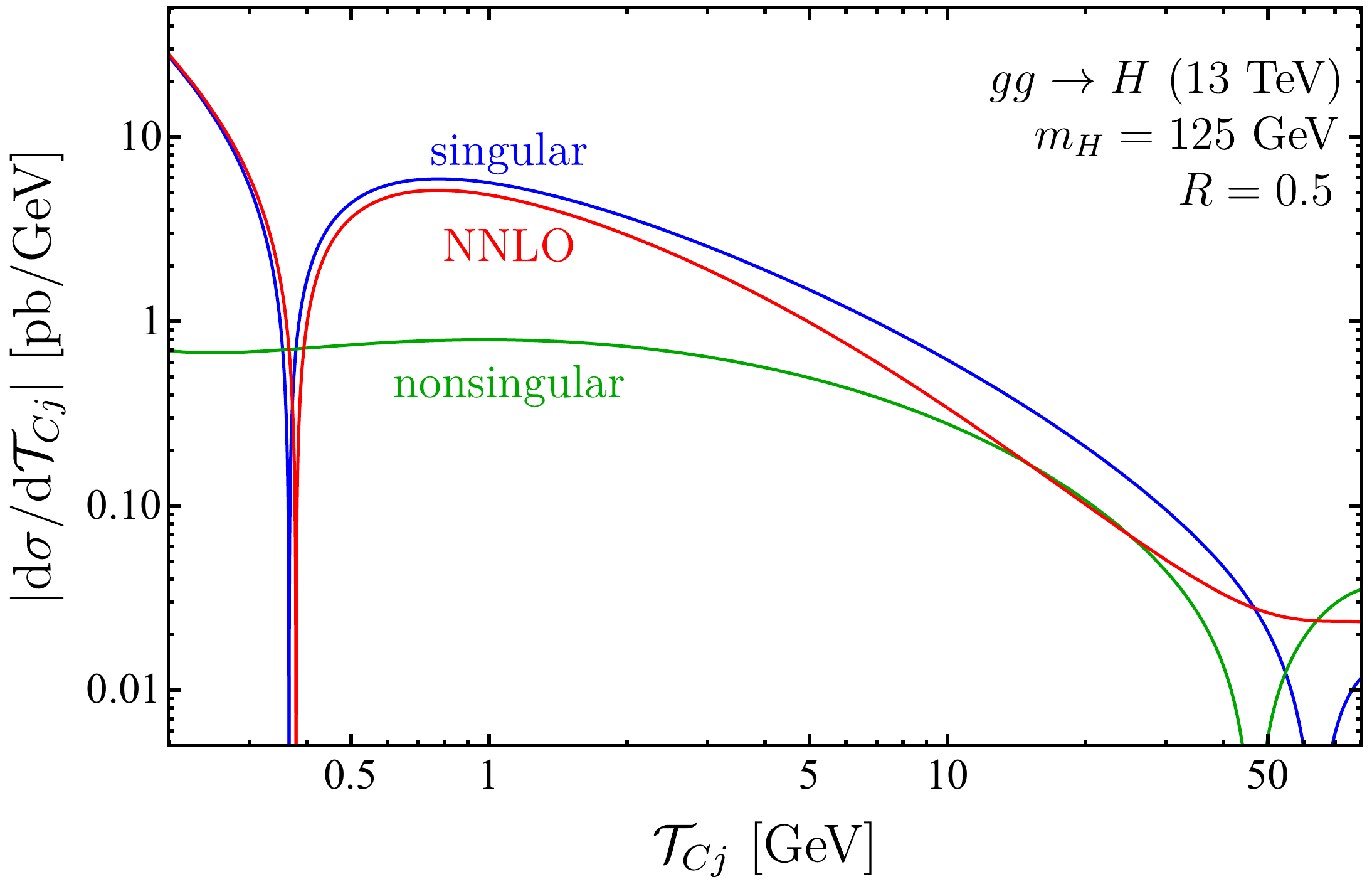}
\caption{The singular, nonsingular and full fixed order NNLO cross section differential in $\Tau_{Bj}$ (left) and $\Tau_{Cj}$ (right) for R=0.5.}
\label{fig:spectrum}
\end{figure}

We plot the singular, nonsingular and total NNLO cross sections differential in $\Tau_{Bj}$ (left) and $\Tau_{Cj}$ (right) in \fig{spectrum}. One observes that at small $\Tau_{fj}$ the computed nonsingular contribution indeed has a rather mild dependence on $\Tau_{fj}$, and does not diverge as strongly as $1/\Tau_{fj}$; this serves as an important cross-check of the two-loop pieces in Eq.~\eqref{eq:TauBCfacto} not proportional to $\delta(\Tau^\cut)$, which also appear in the singular spectrum calculation. Towards larger $\Tau_{fj}$ values, around $\Tau_{fj} \sim 50\,\GeV$, the singular and nonsingular contributions become comparable, and for $\Tau_{fj} \sim 70\,\GeV$ the singular contribution crosses zero. It is interesting to observe that the nonsingular contribution for $\Tau_{Cj}$ is generally larger than that for $\Tau_{Bj}$. 

Comparing \fig{spectrum} to Fig.~2 of Ref.~\cite{Gangal:2014qda}, we see no particular reason to adjust the boundaries of the resummation, transition and fixed-order regions from the ones chosen in the NLL$'$+NLO calculation of Ref.~\cite{Gangal:2014qda}. The two boundaries are chosen to be at $18.75$ GeV and $75$ GeV for $\Tau_{Bj}$, and at $12.5$ GeV and $68.75$ GeV for $\Tau_{Cj}$.

The functional form of the profile scales we use are similar to those adopted in Refs.~\cite{Stewart:2013faa, Gangal:2014qda}. In particular, we have:
\begin{align} \label{eq:centralscale}
\mu_H &= -\img \mu_\FO 
\,, \nn \\
\mu_S(\Tau^\cut) & = \mu_\FO f_\run(\Tau^\cut/m_H)
\,, \nn \\
\mu_B(\Tau^\cut) &= \sqrt{\mu_S(\Tau^\cut) \mu_\FO} = \mu_\FO \sqrt{f_\run(\Tau^\cut/m_H)}
\,, 
\end{align}
where the profile function $f_\run(x)$ is:
\begin{align}
f_{\run}(x) &= 
\begin{cases} x_0 \bigl[1+ (2r_s-1)(x/x_0)^2/4 \bigr] & x \le 2x_0\,,
 \\ r_s x & 2x_0 \le x \le x_1\,,
 \\ r_s x + \frac{(2-r_s x_2- r_s x_3)(x-x_1)^2}{2(x_2-x_1)(x_3-x_1)} & x_1 \le x \le x_2\,,
 \\  1 - \frac{(2-r_s x_1- r_s x_2)(x-x_3)^2}{2(x_3-x_1)(x_3-x_2)} & x_2 \le x \le x_3\,,
 \\ 1 & x_3 \le x\,.
\end{cases}
\label{eq:frun}
\end{align}
For $r_s=1$ the profiles reduce to those of Refs.~\cite{Stewart:2013faa, Gangal:2014qda}. For $2x_0 m_H < \Tau^\cut < x_1 m_H$, $\mu_B$ and $\mu_S$ have canonical scaling: $\mu_S = r_s \Tau^\cut$ and $\mu_B = \sqrt{r_s m_H \Tau^\cut }$ for $\mu_{\mathrm{FO}} = m_H$. The fixed-order region is reached at $\Tau^\cut=x_3m_H$, with a smooth transition being achieved in two steps between $x_1 m_H$ and $x_3 m_H$. We refer to $\Tau^\cut < 2x_0 m_H$ as the `nonperturbative region', and choose $2x_0 m_H$ to be in vicinity of $\Lambda_{QCD}$. In this region the variation of $\mu_{S/B}$ is gradually turned off as $\Tau^\cut \to 0$, with $\mu_S$ and $\mu_B$ approaching the positive values $x_0 \mu_{\mathrm{FO}}$ and $\sqrt{x_0} \mu_{\mathrm{FO}}$ at $\Tau^\cut = 0$. This is to avoid $\alpha_s$ and the PDFs being evaluated at too low scales. In fact our purely perturbative predictions for the cross section will be insufficient in this (small) region, since neglected power corrections can become of $\mathcal{O}(1)$.

For the parameters $x_0 - x_3$ and $\mu_{\mathrm{FO}}$, we make the same choices as in Ref.~\cite{Gangal:2014qda}, which in particular enforces the same boundaries between resummation, transition and fixed-order regions as in that paper (and which are mentioned above). So, for $\Tau_{Bj}$ we have:
\begin{align} \label{eq:TauBprofile}
\mu_\FO &= m_H
\,, \qquad
x_0 = 2.5\GeV/\mu_\FO
\,\,,\,\,
\{x_1,x_2, x_3\} = \{0.15,0.375,0.6\}
\,,\end{align}
and for $\Tau_{Cj}$,
\begin{align} \label{eq:TauCprofile}
\mu_\FO &= m_H
\,, \qquad
x_0 = 2.5\GeV/\mu_\FO
\,, \qquad \nn 
\{x_1,x_2, x_3\} = \{0.1,0.325,0.55\}
\,.\end{align}

The parameter $r_s$ should be chosen to be $\mathcal{O}(1)$. In the next section we will generate results using both $r_s=1$ and $r_s=2$.

\begin{figure}[t]
\includegraphics[width=0.32\textwidth]{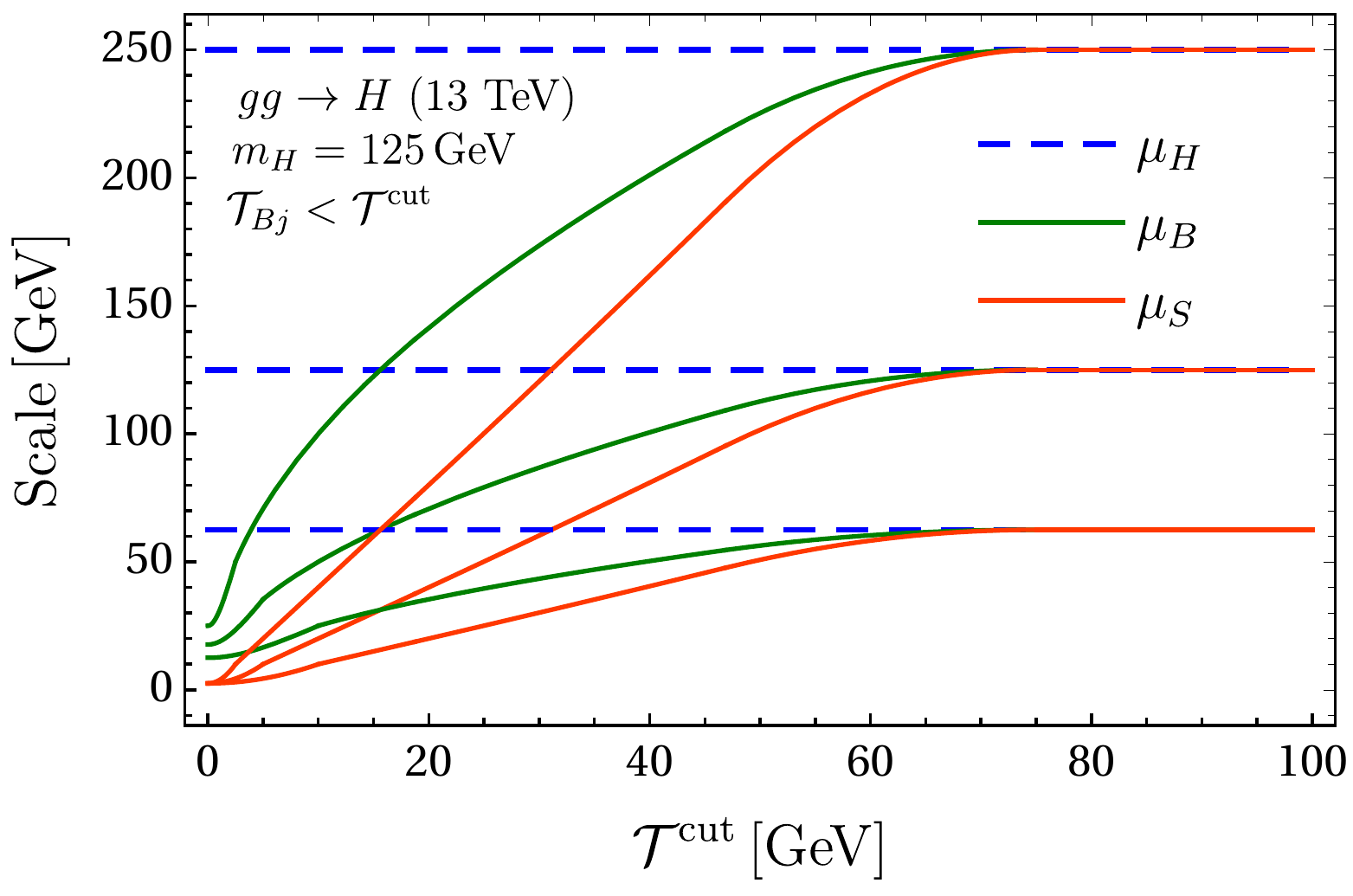}
\hspace{0.005\textwidth}
\includegraphics[width=0.32\textwidth]{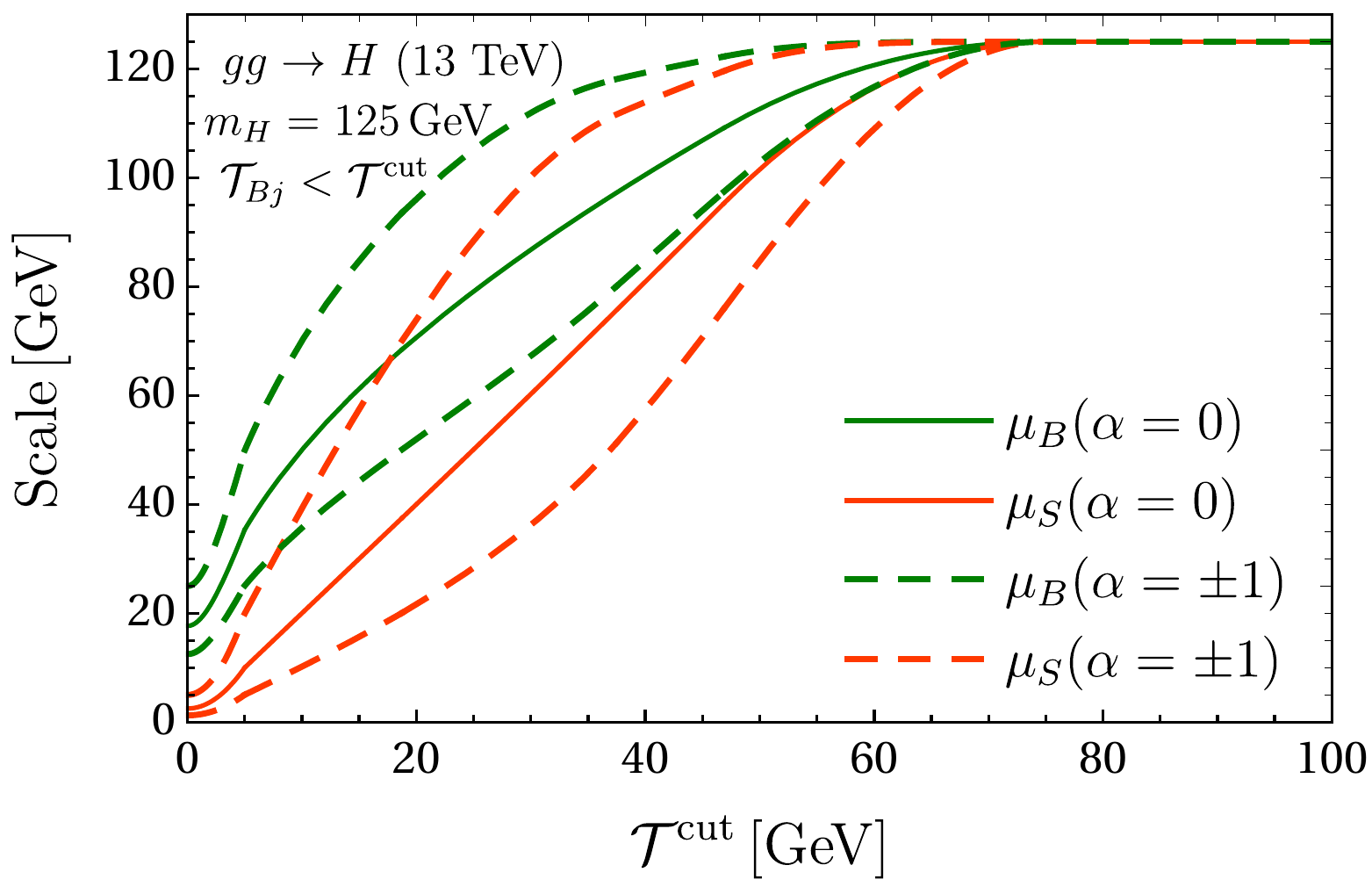}
\hspace{0.005\textwidth}
\includegraphics[width=0.32\textwidth]{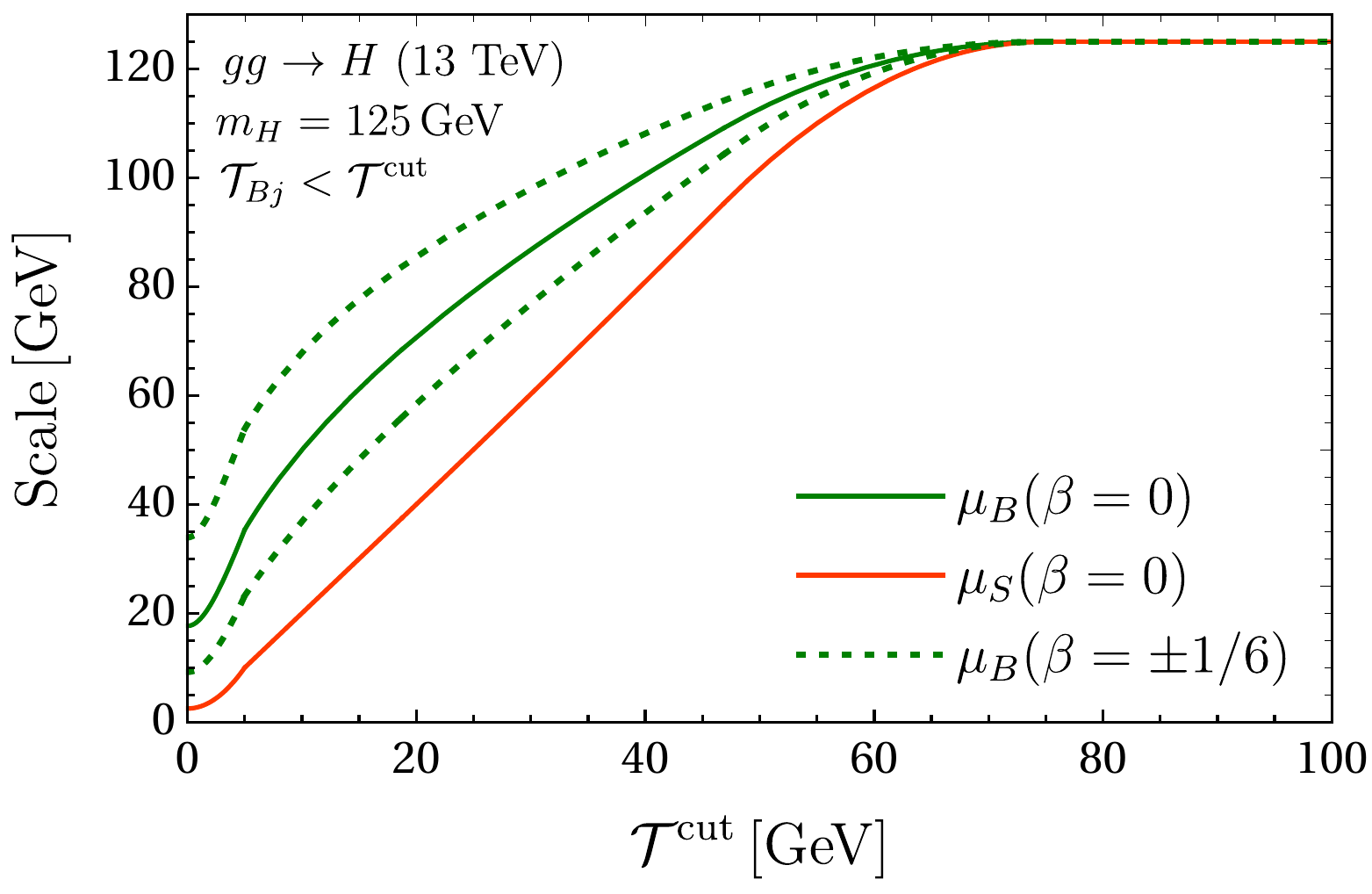}
\vspace{-0.6cm}
\caption{The left panel shows the collective variation of $\mu_B$ and $\mu_S$ scales by a factor of 2 (using $\mu_{\text{FO}} = \{m_H, 2m_H,m_H/2 \}$ in Eq.~\ref{eq:centralscale})) which estimates the fixed order scale uncertainty. The plots in the middle and right panels show $\mu_B$ and $\mu_S$ variation as discussed in the text (Eq.~\ref{eq:muSvary}), used to estimate the resummation uncertainty. All plots in this figure have been generated using $r_s=2$ in Eq.~\eqref{eq:frun}.}
\label{fig:scales}
\end{figure}

Let us now move to a discussion of how we estimate the theoretical uncertainty in our predictions of the jet vetoed cross sections.
These may be parametrized in terms of fully correlated (yield) and fully anti-correlated (migration) components
\cite{Berger:2010xi, Stewart:2011cf, Gangal:2013nxa,Stewart:2013faa}. The yield uncertainty corresponds to the fixed-order uncertainty $\Delta_\mathrm{FO}$. At large $\Tau^\cut$ this reproduces the fixed-order scale variation uncertainty in the total inclusive cross section. The migration uncertainty corresponds to
the uncertainty in the resummed logarithmic series induced by the jet veto cut and is identified as the resummation uncertainty $\Delta_\mathrm{resum}$. The total uncertainty in the
0-jet cross section can be written as
\begin{align}
\Delta_0^2 (\Tau^\cut) = \Delta_\mathrm{FO}^2(\Tau^\cut) + \Delta_\mathrm{resum}^2(\Tau^\cut)\,. 
\end{align}
To estimate these perturbative uncertainties, we vary the profile scales about their central values. For the fixed-order uncertainty $\Delta_\mathrm{FO}$, we vary $\mu_{\text{FO}}$ in the range $\{2m_H, m_H/2\}$ in Eq.~\ref{eq:centralscale}, and the resulting profiles are illustrated for the case $r_s=2$ in the left panel of Fig.~\ref{fig:scales}. The resummation uncertainty $\Delta_\mathrm{resum}$ can be obtained by varying the $\mu_B$ and $\mu_S$ scales using a multiplicative factor,
\begin{align}
f_\vary(x) = \begin{cases}
2(1- (1+\delta) x^2/x_3^2 ) & 0 \le x \le x_3/2
 \\
 1 + 2(1-3\delta)(1-x/x_3)^2 + 16\delta(1-x/x_3)^4 & x_3/2 \le x \le x_3
 \\
1 & x_3 \le x
\end{cases}\,,
\end{align}
For $\delta=0$, this reproduces the functional form of $f_\vary$ used in Ref.~\cite{Gangal:2014qda}. Here we set $\delta=0.05$, so that when using $r_s=2$, we avoid either $\mu_B$ or $\mu_S$ rising above the fixed-order value of $m_H$ in any of the variations discussed below.

The up and down variations of $\mu_B$ and $\mu_S$ are parametrized using this multiplicative factor as follows,
\begin{align} \label{eq:muSvary}
\mu_S^\vary(x, \alpha) &= f_\vary^\alpha (x)\, \mu_S(x)  =  \mu_\FO\, f_\vary^\alpha (x) \,f_\run(x)\,, \nonumber\\
\mu_B^\vary(x, \alpha,\beta) &= {\mu_S^\vary(x, \alpha)}^{1/2-\beta} \mu_\FO^{1/2+\beta}
= \mu_\FO \bigl[f_\vary^\alpha(x)\, f_\run(x) \bigr]^{1/2-\beta}
\,,\end{align}
with $(\alpha, \beta) = \{(+1,0), (-1,0), (0,+1/6), (0,-1/6)\}$. The variations in the parameter $\alpha$ lead to a factor of 2 variation in $\mu_S$ for $\Tau^\cut \to 0$,
and a factor of $\sqrt{2}$ variation in $\mu_B$. The profiles thus obtained with $\alpha$ variation are shown for $r_s=2$ by the dashed curves in the middle panel of Fig.~\ref{fig:scales}.
The parameter $\beta$ modifies $\mu_B$ by varying the canonical relation $\mu_B \sim \sqrt{\mu_S \mu_H}$ while keeping $\mu_S$ fixed,
as is shown for $r_s=2$ by the dotted lines in the right panel of Fig.~\ref{fig:scales}. A detailed discussion about the choice of these parameters can be found in Ref.~\cite{Gangal:2014qda}. These variations
in $\mu_S$ and
$\mu_B$ vary the arguments of the logarithms in the evolution kernel, and thus help in estimating the higher-order corrections in the resummed results. The total resummation uncertainty $\Delta_\mathrm{resum}$
is obtained by taking the maximum of the absolute deviation from the central profile. 

\section{Resummed Predictions at NNLL$'+$NNLO }
\label{sec:III}

In this section we present our results for the $0$-jet gluon-fusion Higgs cross section at NNLL$'+$ NNLO with the jet veto imposed via the $\Tau_{Bj}$ and $\Tau_{Cj}$ variables. We compare these results to NLL$'+$NLO and NLL results for the same cross sections, to study the perturbative convergence.

Let us briefly summarise our set-up. For the PDFs we use the MMHT sets \cite{Harland-Lang:2014zoa}: the mmht2014lo135 set for the NLL, the mmht2014nlo120 set for the NLL$'+$NLO, and the mmht2014nnlo118 set for the NNLL$'+$NNLO predictions. We take $m_H = 125$ GeV, $m_t=172.5$ GeV, $m_b=4.7$ GeV, and set $n_f = 5$ in all perturbative ingredients. The effect of finite $m_b, m_t$ are taken into account in the hard function of the resummation $H_{gg}$ and the nonsingular cross section up to NLO; for the NNLO coefficient of the nonsingular cross section we use the $m_t\to\infty,m_b\to 0$ result, whereas for the NNLO coefficient of $H_{gg}$ we use the approach described under Eq.~\eqref{eq:Hfuncexp}.  The central prediction has $i\mu_H = \mu_{\text{FO}} = m_H$ and $\mu_B, \mu_S$ defined by the profile functions in Eq.~\eqref{eq:centralscale}. The uncertainty bands are estimated using profile scale variations as discussed in Sec.~\ref{sec:II}.

\begin{figure}[t]
\includegraphics[width=0.46\textwidth]{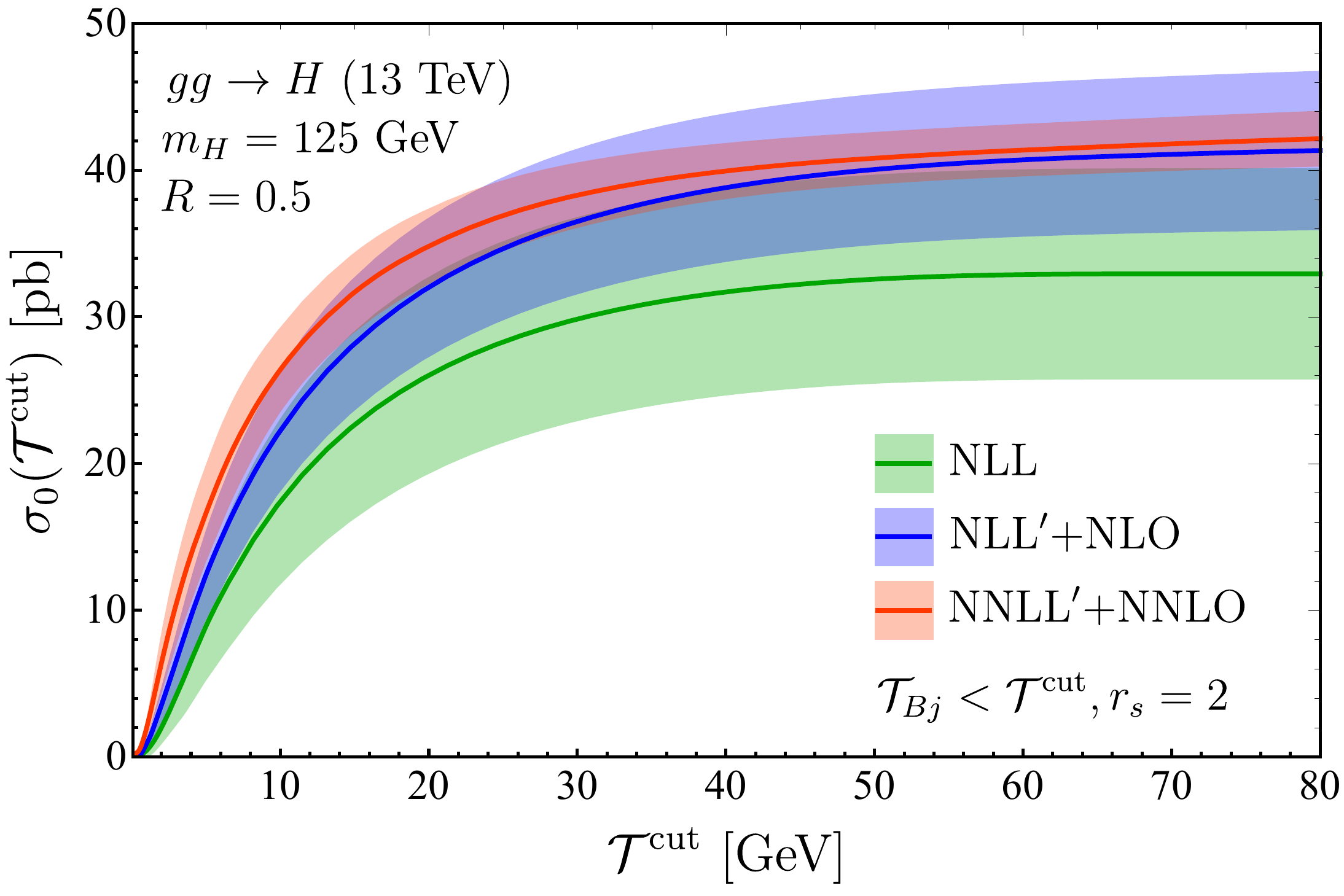}
\hspace{0.02\textwidth}
\includegraphics[width=0.46\textwidth]{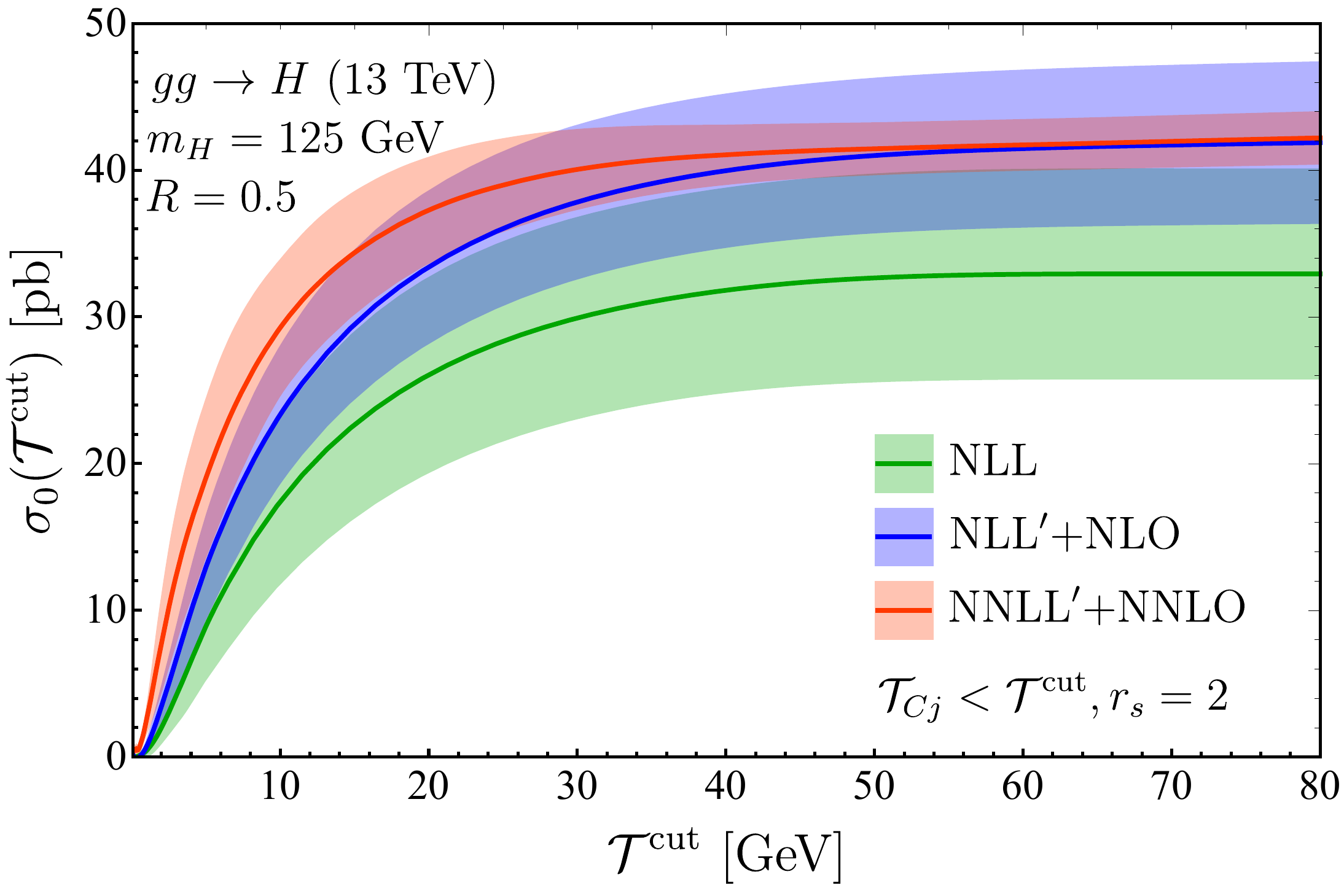}
\caption{The dark orange band shows the cumulant NNLL$'+$NNLO cross section for $\Tau_{Bj} < \Tau^\cut$ (left panel) and $\Tau_{Cj} < \Tau^\cut$ (right panel) for $R = 0.5$. The blue and green bands correspond to NLL$'+$NLO and NLL predictions respectively, for each of the two observables. The solid lines indicate the predictions using the central values of the profile scales. These results have been obtained using $r_s=2$ in Eq.~\eqref{eq:frun}.}
\label{fig:NNLL}
\vspace{2mm}
\end{figure}

\begin{figure}[t]
\includegraphics[width=0.46\textwidth]{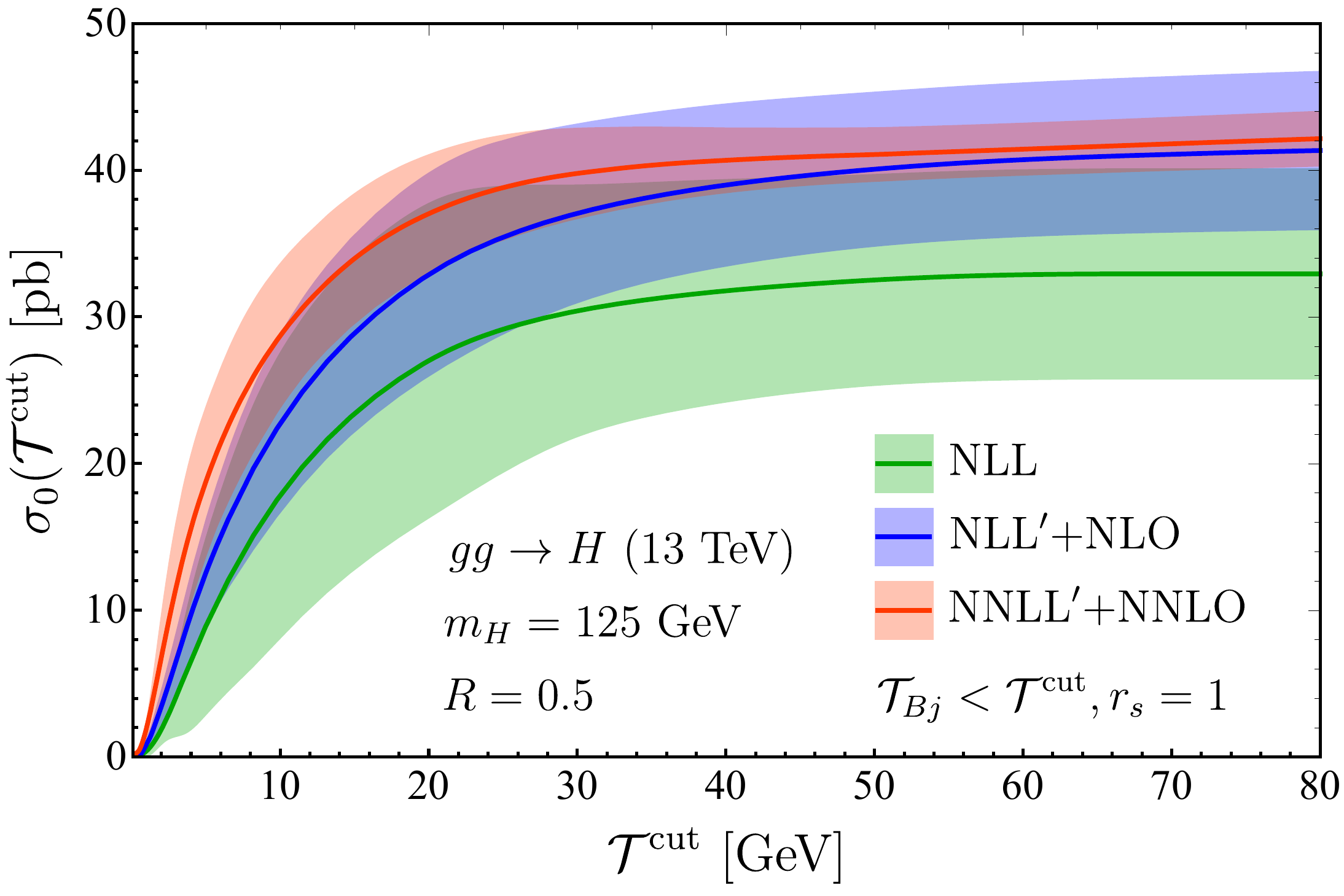}
\hspace{0.02\textwidth}
\includegraphics[width=0.46\textwidth]{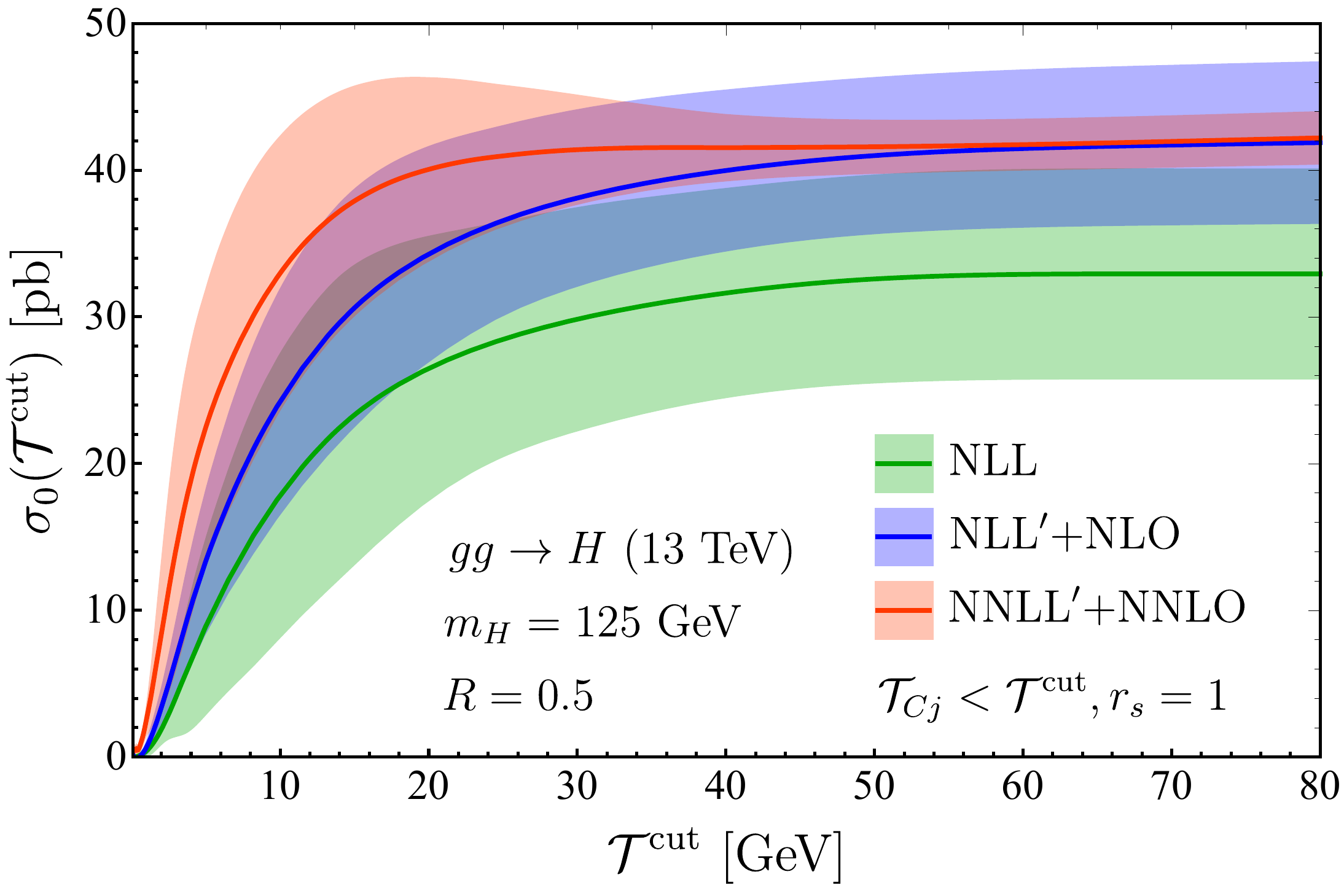}
\caption{The same plots as in \fig{NNLL}, but now generated setting $r_s=1$ in Eq.~\eqref{eq:frun}.}
\label{fig:NNLLrs1}
\vspace{2mm}
\end{figure}

\begin{figure}[t]
\includegraphics[width=0.455\textwidth]{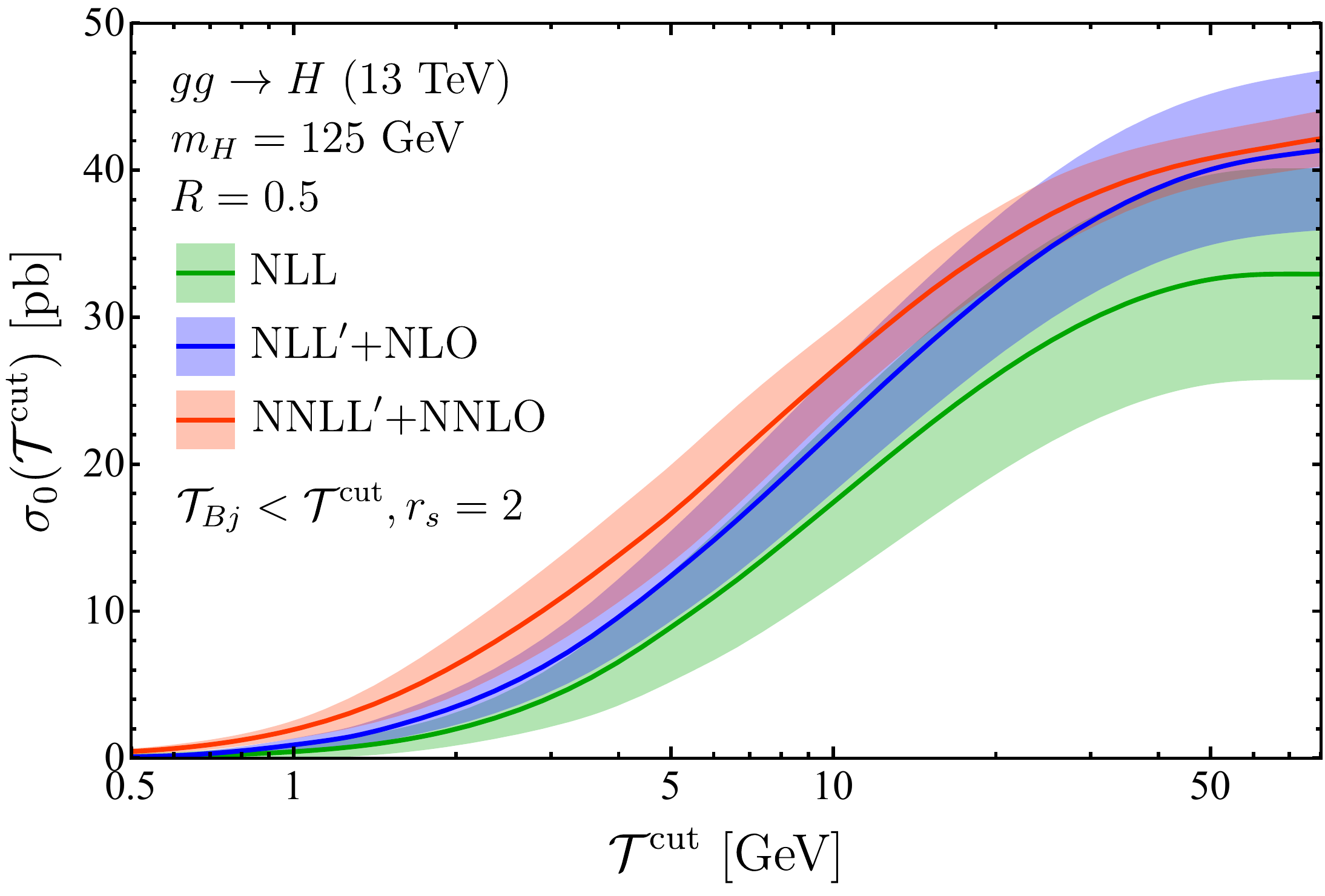}
\hspace{0.02\textwidth}
\includegraphics[width=0.455\textwidth]{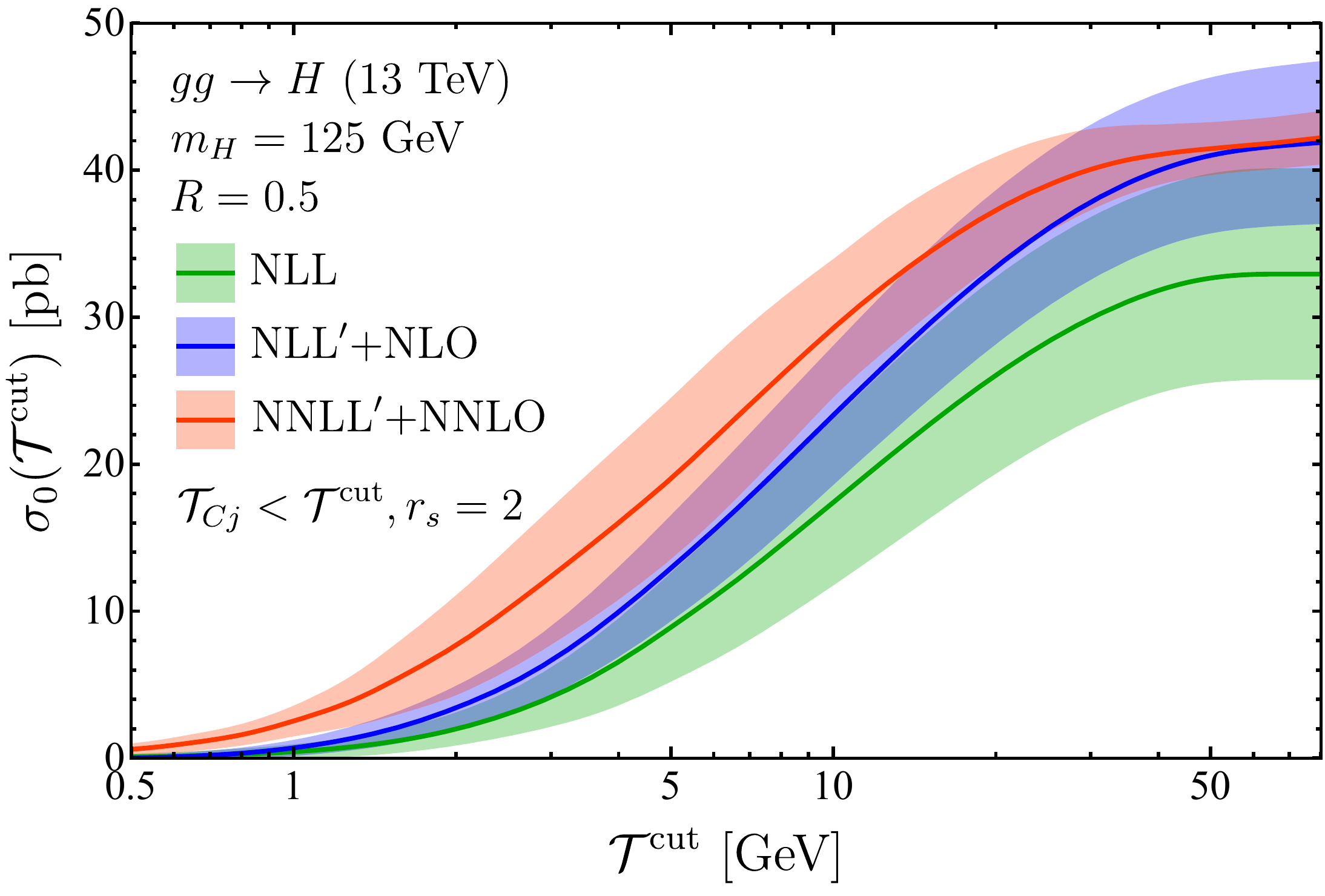}
\caption{The same plots as in \fig{NNLL}, but with a logarithmic $x$-axis.}
\label{fig:NNLLlog}
\vspace{2mm}
\end{figure}

\begin{figure}[t]
\includegraphics[width=0.455\textwidth]{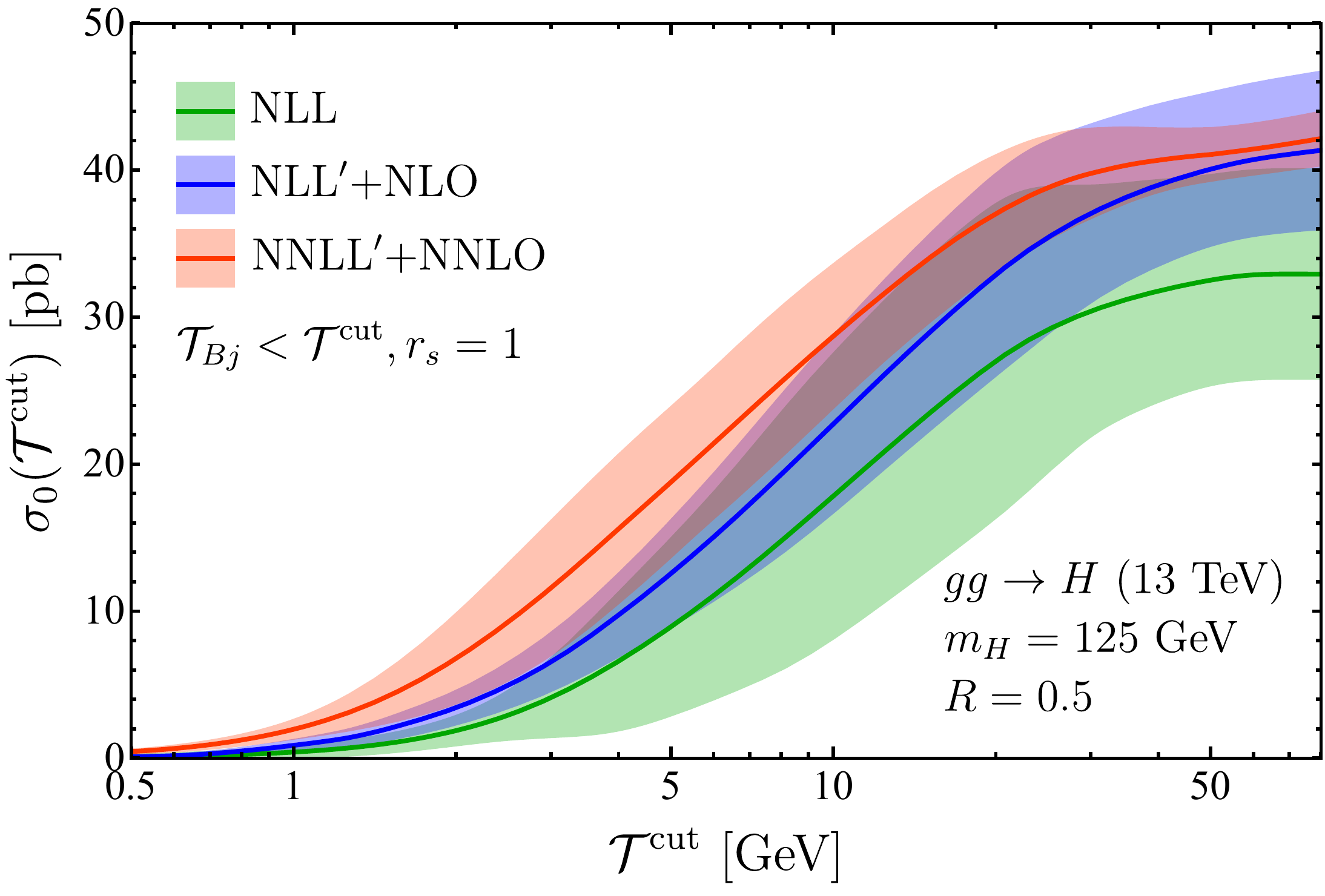}
\hspace{0.02\textwidth}
\includegraphics[width=0.455\textwidth]{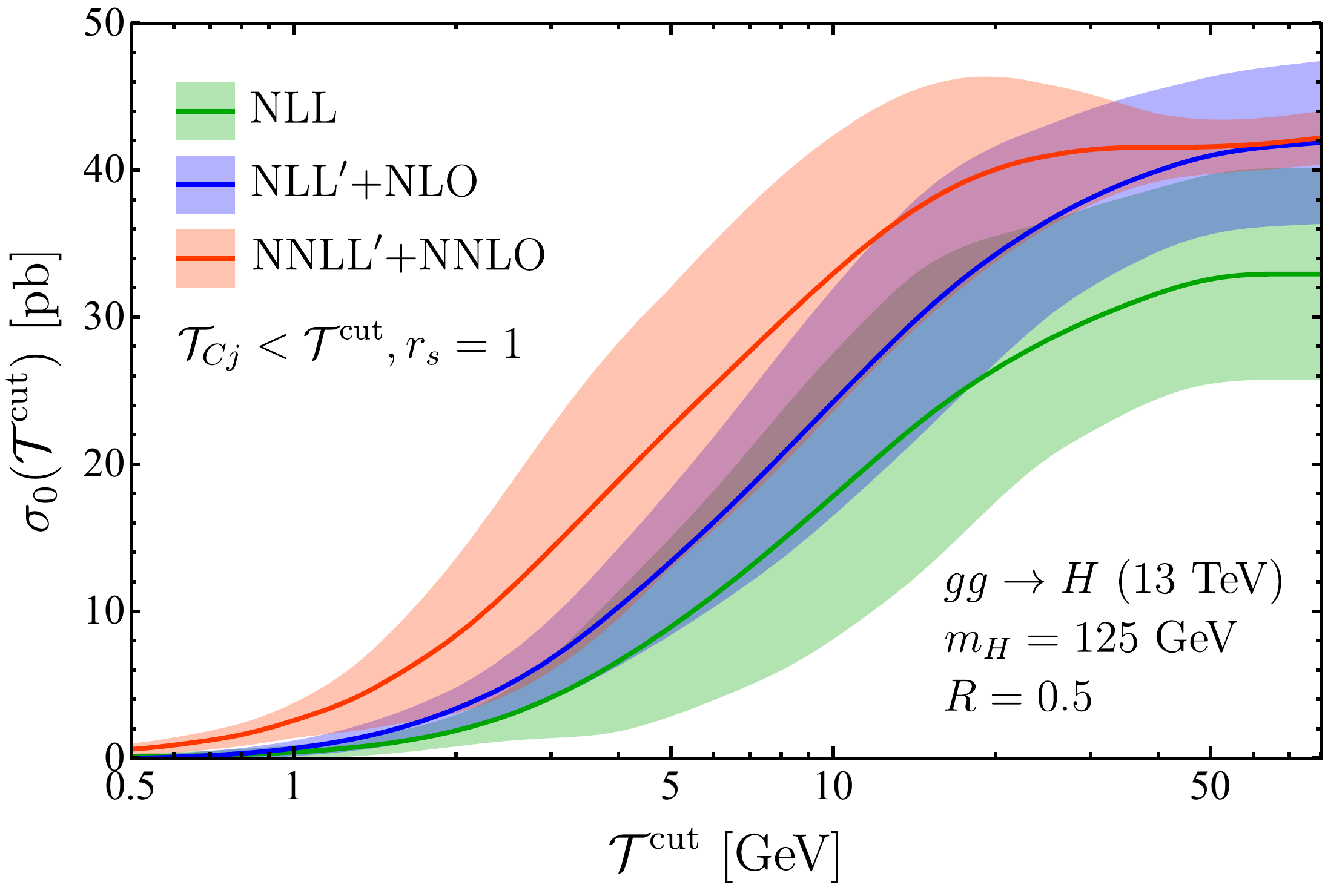}
\caption{The same plots as in \fig{NNLLrs1}, but with a logarithmic $x$-axis.}
\label{fig:NNLLlogrs1}
\end{figure}

We plot the cross section with $\Tau_{B/Cj} < \Tau^\cut$ as a function of $\Tau^\cut$, for two choices of $r_s$ in Eq.~\eqref{eq:frun}: $r_s=1$ and $r_s=2$. In each case we make plots using both a linear and a logarithmic $x$-axis. The results for $r_s=2$ are given in \fig{NNLL} and \fig{NNLLlog}, and those for $r_s=1$ are given in \fig{NNLLrs1} and \fig{NNLLlogrs1}. In each figure, we plot our
NNLL$'+$NNLO results in orange, and give NLL$'+$NLO and NLL results in blue and green respectively.

In general, we see a substantial reduction of uncertainties going from NLL$'$ to NNLL$'$, due to the increase in the accuracy of resummation as well as matching. The predictions of higher orders fall within the uncertainty bands of lower orders, thereby indicating a good perturbative convergence.

Comparing the results with $r_s=2$ to those with $r_s=1$, we see a better convergence between different orders in resummed perturbation theory in the former case, where this is particularly noticeable for $\Tau_{Cj}$. In the case of $\Tau_{Cj}$ with $r_s=1$, one also observes that the top of the uncertainty band at $\Tau^\cut \sim 20$ GeV greatly exceeds that of the total NNLO cross section. This is of course unphysical. For these reasons we would advocate to use the results with $r_s=2$ for $\Tau_{Cj}$. The choice of $r_s=2$, corresponding to the choice of canonical scale $\mu_S = 2\Tau^\cut$, also seems physically reasonable for $\Tau_{Cj}$, since the cut $\Tau_{Cj}<\Tau^\cut$ corresponds to an $m_{Tj}$ cut that is everywhere looser than $m_{Tj}<2\Tau^\cut$. For the case of $\Tau_{Bj}$, there is not a strong difference between the $r_s=1$ or $r_s=2$ results and one can use either of these. 

We note in passing that similar conclusions with regards to the scales in the resummation region were also found for thrust and $C$-parameter in $e^+e^-$ collisions, in Ref.~\cite{Hoang:2014wka}. Furthermore, we note that in Ref.~\cite{Berger:2010xi}, where NNLL+NNLO predictions were obtained for Higgs production with a cut on beam thrust (the global equivalent of $\Tau_{Bj}$), profile scales that are rather close to $\mu_S = 2\Tau^\cut, \mu_B = \sqrt{m_H\mu_S}$ in the resummation region were used.

Finally, in Table~\ref{table-0jetresult} we present numbers for the 
$H$+0-jet cross section defined by $\Tau_{B/Cj} < \Tau^\cut$, at two sample values of $\Tau^\cut$: 20 GeV and 30 GeV. We give the central predictions and perturbative uncertainties, for both the $r_s=1$ and $r_s=2$ cases.

These numbers highlight the improvement in perturbative convergence going from $r_s=1$ to $r_s=2$ that was already visible in Figs.~\ref{fig:NNLL} - \ref{fig:NNLLlogrs1}. For the results with $r_s=2$, the perturbative uncertainties are smaller compared to the results with $r_s=1$, and the NNLL$'$+NNLO results lie closer to the NLL$'+$NLO ones for $r_s=2$. We again see that this is more pronounced for the $\Tau_{Cj}$ case, leading to the choice $r_s=1$ being disfavoured for $\Tau_{Cj}$.

\begin{table}[h]
  \begin{center}
\begin{tabular}{|c|c|c|}
\hline
   & \quad \quad\ $\sigma_0(\Tau^\cut) {\rm [pb]}\, (r_s =1)$ \quad\quad &\quad\quad $\sigma_0(\Tau^\cut){\rm [pb]}\, (r_s = 2)$ \quad \quad  \\
\hline 
\quad \quad NLL$'+$NLO \quad \quad & &   \\
\quad \quad $\Tau_{Bj} < \Tau^\cut = 20\, \rm GeV$ \quad \quad  & \quad \quad $32.88 \pm 6.95 \, (21.2 \%)$ \quad \quad & \quad \quad $32.02 \pm 4.75 \, (14.8 \%)$ \quad \quad \\ 
\quad \quad $\Tau_{Bj} < \Tau^\cut = 30\, \rm GeV$ \quad \quad  & \quad \quad $37.05 \pm 6.12 \, (16.5 \%) $ \quad \quad  &  \quad \quad $36.50 \pm 4.96 \, (13.6 \%) $ \quad \quad \\
\hline
\quad \quad NNLL$'+$NNLO \quad \quad & &  \\
\quad \quad $\Tau_{Bj} < \Tau^\cut = 20\, \rm GeV$ \quad \quad  & \quad \quad $37.03 \pm 4.06 \, (10.9 \%)$ \quad \quad & \quad \quad $34.81 \pm 2.57 \, (7.39 \%)$ \quad \quad \\ 
\quad \quad $\Tau_{Bj} < \Tau^\cut = 30\, \rm GeV$ \quad \quad & \quad \quad $39.77 \pm 3.11 \, (7.82\%) $ \quad \quad  &  \quad \quad 
$ 38.30 \pm 2.23 (5.82 \%) $ \quad \quad \\
\hline \hline
\quad \quad NLL$'+$NLO \quad \quad & & \\
\quad \quad $\Tau_{Cj} < \Tau^\cut = 20\, \rm GeV$ \quad \quad  & \quad \quad $34.28 \pm 7.37 \, (21.5 \%)$ \quad \quad & \quad \quad $33.40 \pm 5.24 \, (15.7 \%)$ \quad \quad \\ 
\quad \quad $\Tau_{Cj} < \Tau^\cut = 30\, \rm GeV$ \quad \quad & \quad \quad $ 38.10 \pm 6.05 \, (15.8 \%)$ \quad \quad & \quad \quad $37.82 \pm 5.27 \, (13.9 \%)$ \quad \quad \\
\hline
\quad \quad NNLL$'+$NNLO \quad \quad & &  \\
\quad \quad $\Tau_{Cj} < \Tau^\cut = 20\, \rm GeV$ \quad \quad  & \quad \quad $40.05 \pm 6.28 \, (15.69 \%)$ \quad \quad & \quad \quad $37.27 \pm 3.64 \, (9.77 \%)$ \quad \quad \\ 
\quad \quad $\Tau_{Cj} < \Tau^\cut = 30\, \rm GeV$ \quad \quad & \quad \quad $ 41.39 \pm 3.75 \, (9.07 \%)$ \quad \quad & \quad \quad $40.05 \pm 2.75 \, (6.88 \%)$ \quad \quad \\
\hline \hline
    \end{tabular}
        \caption{Predictions for the $\Tau_{Bj}$ and $\Tau_{Cj}$ $H+0$-jet cross sections obtained using the central profile scales, along with the total perturbative uncertainties. The equivalent percentage uncertainties are shown in brackets.}
 \label{table-0jetresult}
  \end{center}
\end{table}

\section{Effect of underlying event and hadronisation}
\label{sec:IV}

The analytic resummed calculations presented in the previous section do not take into account the effects of underlying event or hadronisation. These effects are formally suppressed, but it is interesting to assess their practical numerical impact, and compare how much $\Tau_{B/Cj}$ are affected with respect to the conventional jet veto observable $p_{Tj}$.

In order to do this, we employ a NLO + parton shower set-up, as implemented in {\sc MadGraph5\_aMC@NLO} interfaced with Pythia8. In Pythia8, the parameters have been set to their default values as specified by the {\sc MadGraph5\_aMC@NLO} interface. These parameter values mainly correspond to the Monash 2013 tune \cite{Skands:2014pea} -- in particular the underlying event and hadronisation parameters are set as in this tune. We generate ggF Higgs events, compute the cumulant distributions for $\Tau_{Bj}$, $\Tau_{Cj}$ and $p_{Tj}$ from these events, and investigate the effect of turning on and off hadronisation and underlying event (UE) in the shower on these cumulant distributions. Jets with $R=0.5$ are identified in the events using the anti-$k_T$ algorithm \cite{Cacciari:2008gp}.

\begin{figure}[t]
\centering
\hspace{-0.5cm}
\includegraphics[width=0.46\textwidth]{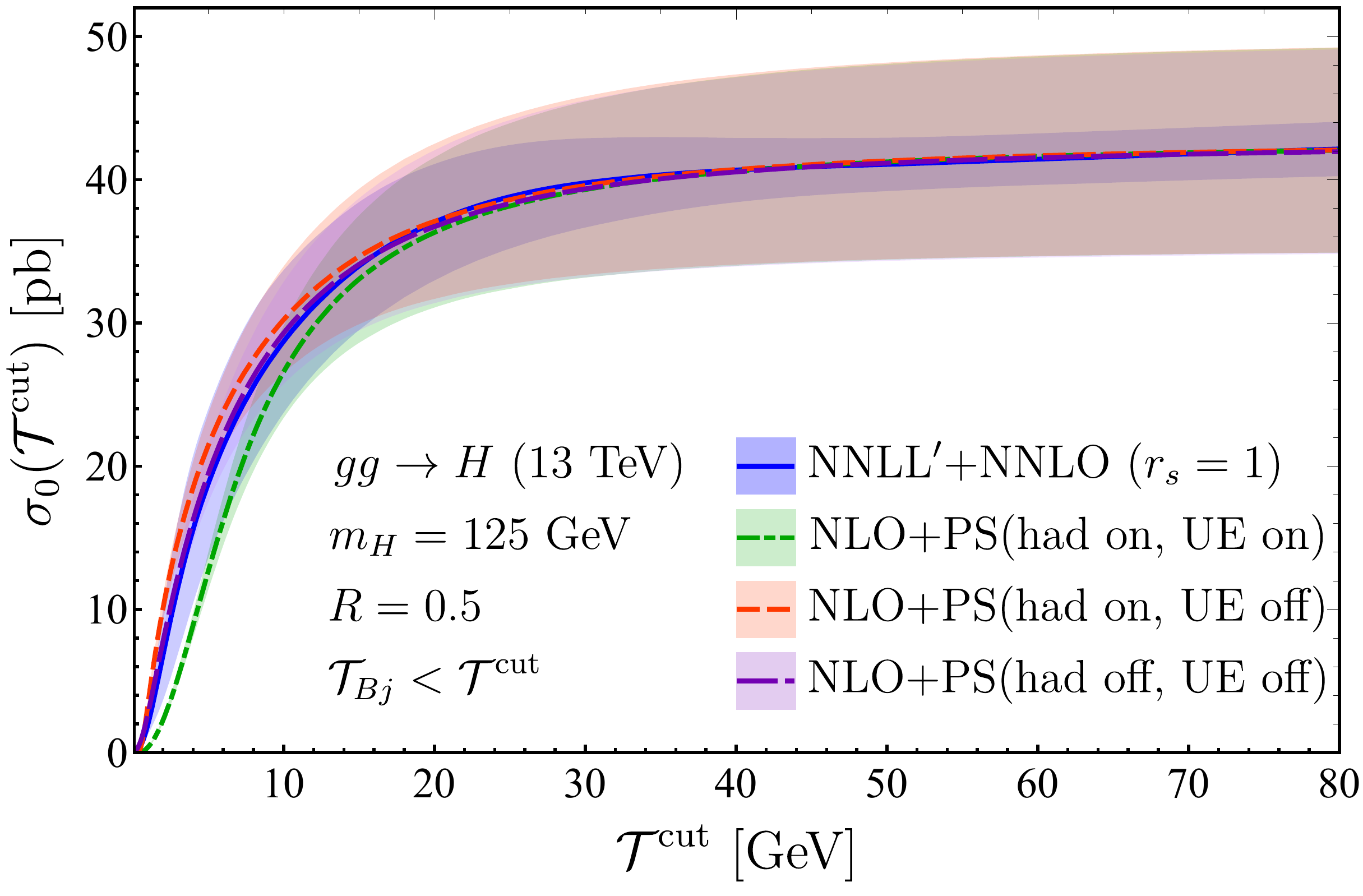}
\hspace{0.02\textwidth}
\includegraphics[width=0.46\textwidth]{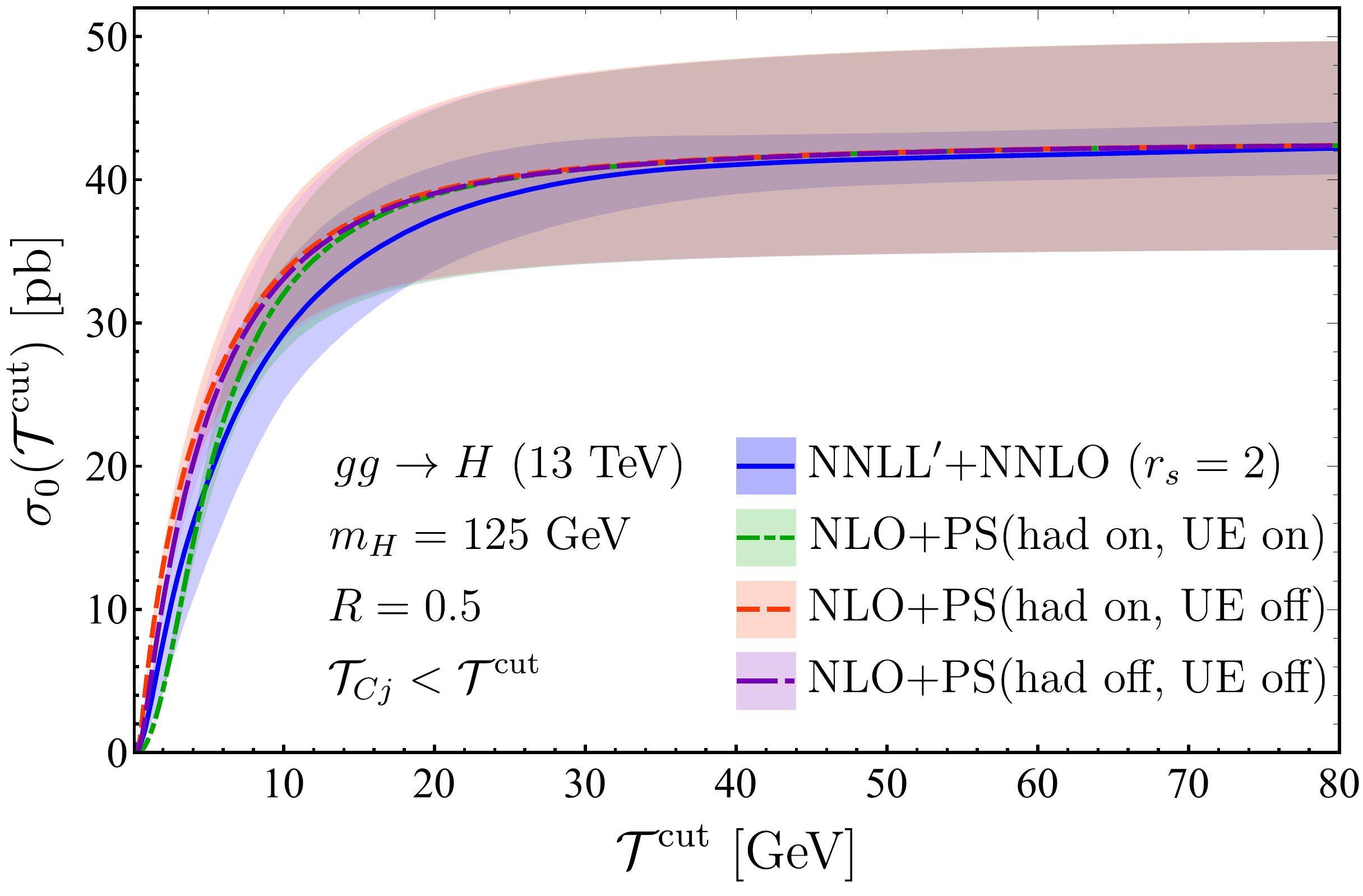}
\includegraphics[width=0.46\textwidth]{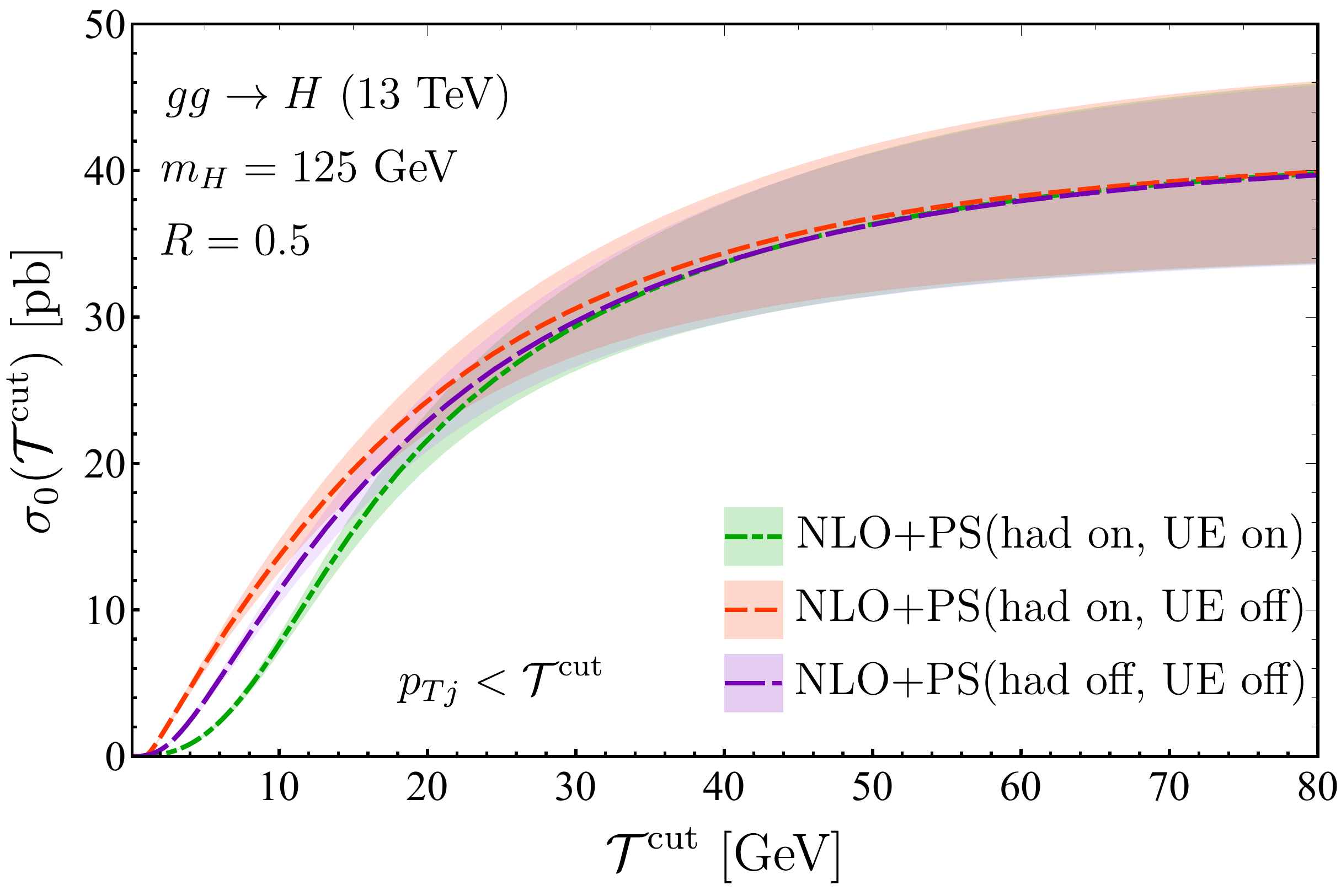}
\caption{NLO + parton shower results for the Higgs cross section for $\Tau_{B/Cj},p_{Tj} < \Tau^\cut$ and $R=0.5$, generated using {\sc MadGraph5\_aMC@NLO} interfaced with Pythia8. The overall normalisation of these predictions has been adjusted such that the central predictions reproduce the NNLO Higgs cross section (including resummation of time-like logarithms) for $\Tau^\cut \to \infty$. The green, red and purple bands represent various configurations with regards to the hadronisation and underlying event. For the $\Tau_{B/Cj}$ cases, we also plot in blue the NNLL$'+$NNLO  analytic prediction for comparison.}
\label{fig:NLOPS}
\end{figure}

\begin{figure}[t]
\centering
\hspace{-0.5cm}
\includegraphics[width=0.46\textwidth]{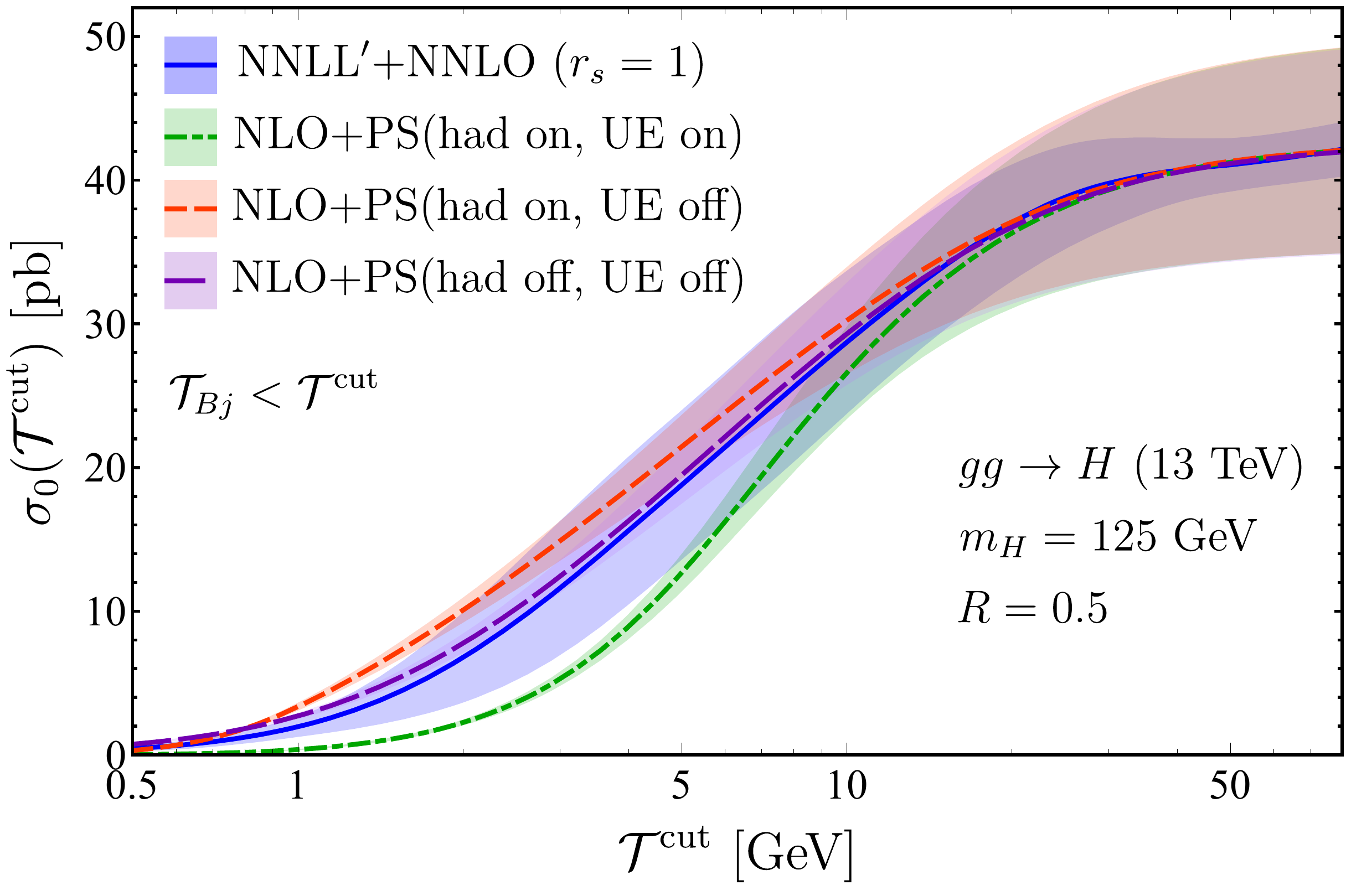}
\hspace{0.02\textwidth}
\includegraphics[width=0.46\textwidth]{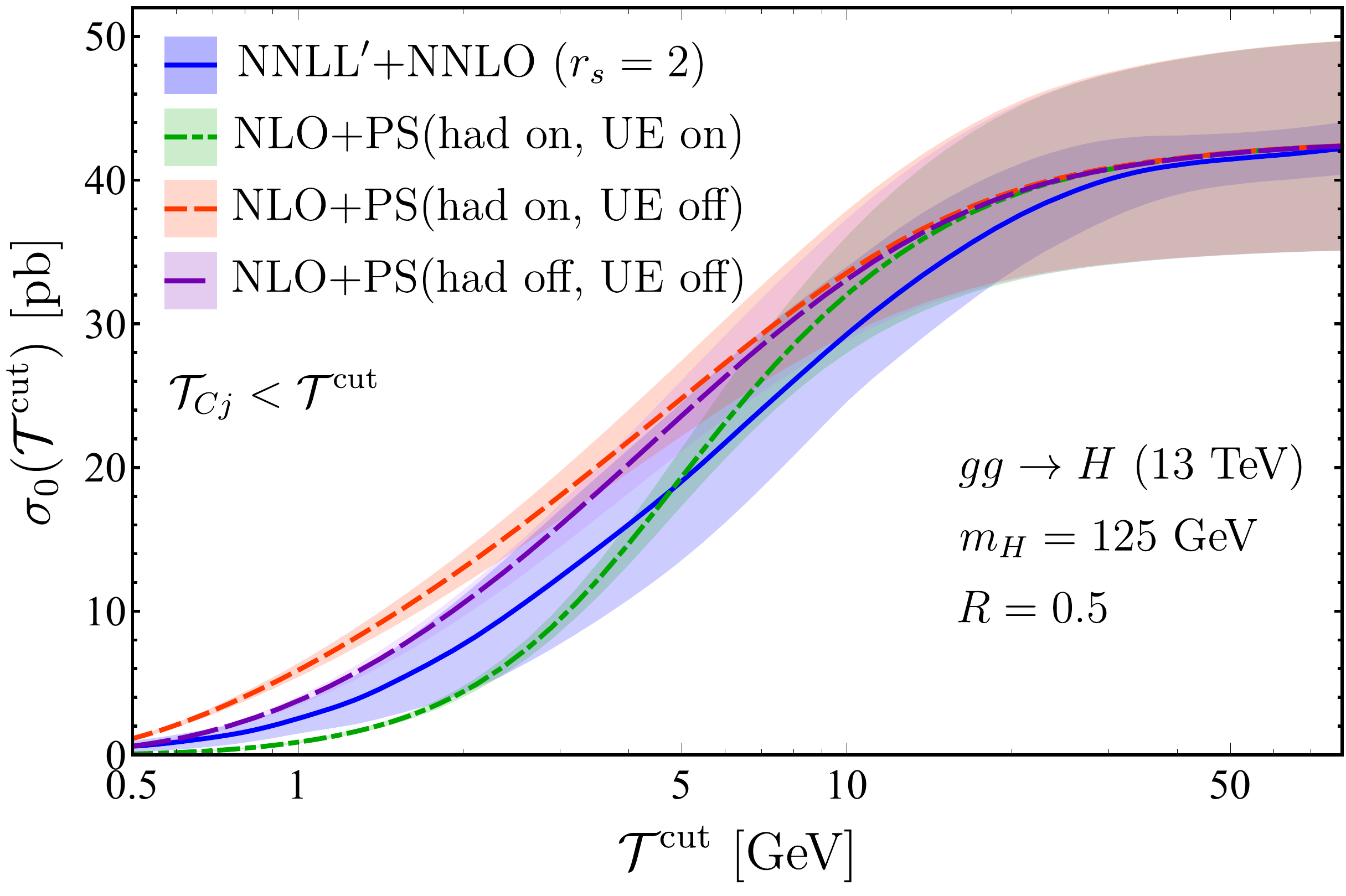}
\includegraphics[width=0.46\textwidth]{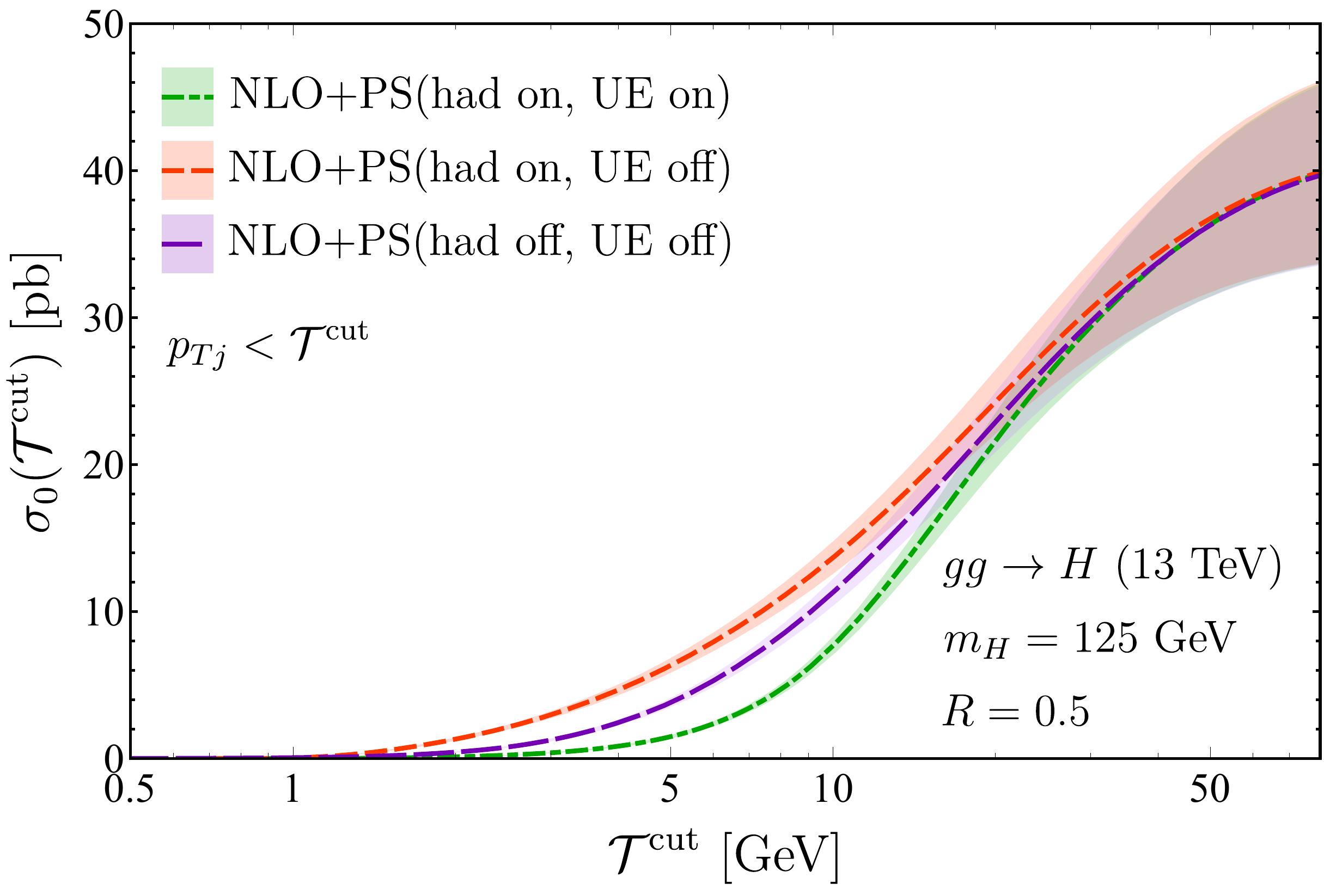}
\caption{The same plots as in \fig{NLOPS}, but with a logarithmic $x$-axis.}
\label{fig:NLOPSlog}
\end{figure}

In \fig{NLOPS}, we give the NLO+PS plots for all jet vetoes, for three configurations of the parton shower: hadronisation and UE turned on, hadronisation on and UE off, and both hadronisation and UE turned off. \fig{NLOPSlog} presents the same plots, but with a logarithmic $x$-axis. For the $\Tau_{B/Cj}$ cases, we also give the NNLL$'+$NNLO prediction, using the same set-up as in section \ref{sec:III} and taking $r_s=1$ ($r_s=2$) for $\Tau_{Bj}$ ($\Tau_{Cj}$). To allow for a meaningful comparison, we rescale all NLO+PS predictions to the total cross section of our NNLL$'+$NNLO predictions. The uncertainty bands on the NLO+PS results reflect the (rescaled) fixed-order scale variation.

From these plots, one can see that hadronisation tends to shift the $p_{Tj}$ and $\Tau_{B/Cj}$ distributions towards lower values, whilst UE pushes these distributions towards higher values. Non-perturbative hadronisation `smears out' the energy in QCD particle sprays over a larger area in $\eta - \phi$ phase space, resulting in a loss of $p_{Tj}$ and $\Tau_{B/Cj}$ from any given jet with fixed $R$. UE sprays extra particles fairly evenly over the $\eta - \phi$ phase space, with these extra particles pushing up the $p_{Tj}$ and $\Tau_{B/Cj}$ values of all jets \cite{Dasgupta:2007wa}.

One can also already see from these plots that the cross section with a $p_{Tj}$ veto is more strongly affected by hadronisation and UE effects than those with $\Tau_{Bj}$ and $\Tau_{Cj}$ vetoes. Finally, it is interesting to note that the rescaled NLO+PS predictions for $\Tau_{Bj}$ and $\Tau_{Cj}$ turn out to lie fairly close to the NNLL$'+$NNLO predictions.

\begin{figure}[t]
\includegraphics[width=0.46\textwidth]{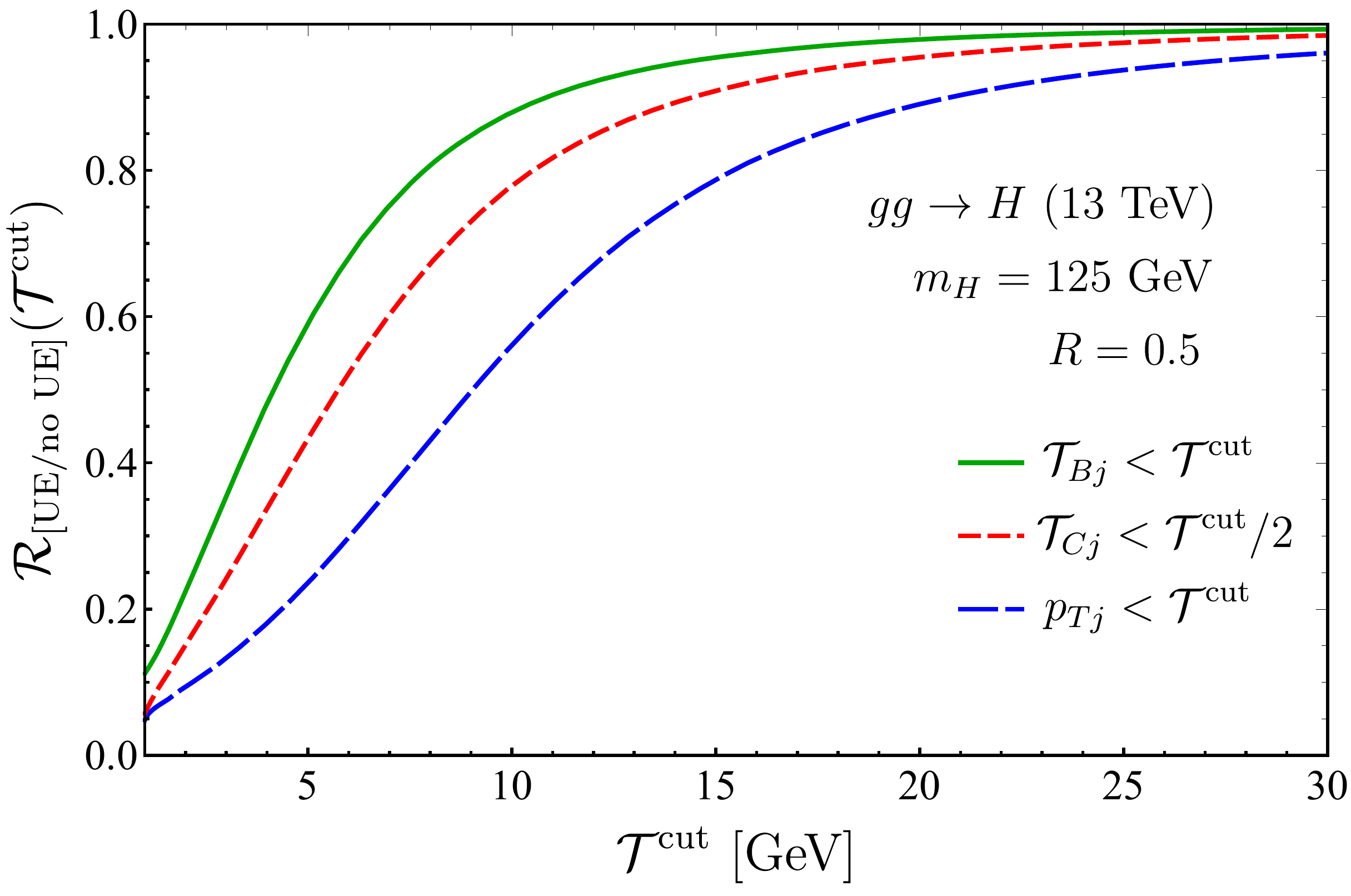}
\hspace{0.02\textwidth}
\includegraphics[width=0.46\textwidth]{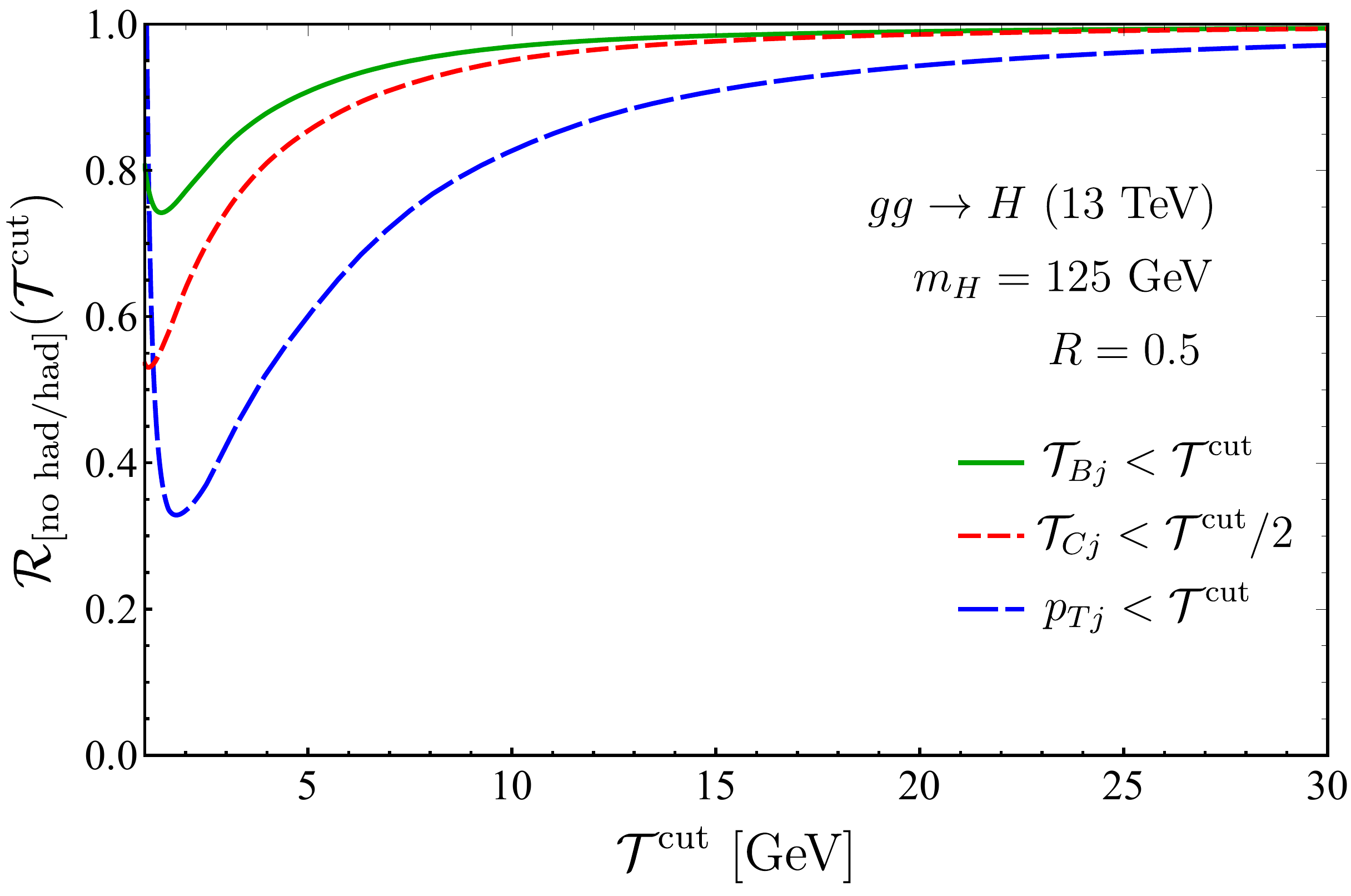}
\caption{Plots for the ratios $\mathcal{R}_{(\text{UE}/\text{no UE})}(\Tau^\cut)$ and $\mathcal{R}_{(\text{no had}/\text{had})}(\Tau^\cut)$ defined in Eqs.~\eqref{eq:UEratio} and \eqref{eq:hadratio}.}
\label{fig:ratios}
\end{figure}

In order to exhibit in more detail the extent to which all three vetoed cross sections are sensitive to UE and hadronisation, we plot in \fig{ratios} the following ratios:
\begin{align} \label{eq:UEratio}
\mathcal{R}_{(\text{UE}/\text{no UE})}(\Tau^\cut) = \dfrac{\sigma_0\left(\Tau^\cut\right)|_{\text{had on, UE on}}}{\sigma_0\left(\Tau^\cut\right)|_{\text{had on, UE off}}}
\\ \label{eq:hadratio}
\mathcal{R}_{(\text{no had}/\text{had})}(\Tau^\cut) = \dfrac{\sigma_0\left(\Tau^\cut\right)|_{\text{had off, UE off}}}{\sigma_0\left(\Tau^\cut\right)|_{\text{had on, UE off}}}
\end{align}
The first ratio indicates the extent to which each observable is sensitive to UE, and the second indicates the extent to which each observable is sensitive to hadronisation. The closer the $\mathcal{R}$ value is to $1$, the lower the sensitivity is. For $p_{Tj}$ and $\Tau_{Bj}$, the cumulant cross sections $\sigma_0\left(\Tau^\cut\right)$ are computed by integrating $p_{Tj}$ or $\Tau_{Bj}$ up to $\Tau^\cut$ as before, whilst for $\Tau_{Cj}$ we integrate up to $\Tau^\cut/2$. We do this so that at a given point on the $x$-axis in \fig{ratios} all the different veto observables correspond to the same `central' $p_{Tj}$ veto at $y_j=Y$ in the limit of small jet radius $R$ (namely, $p_{Tj}|_{y_j=Y,R\ll1}<\Tau^\cut$).

These results confirm that the cross sections with $\Tau_{Bj}$ and $\Tau_{Cj}$ vetoes are less sensitive to UE and hadronisation effects than that with the $p_{Tj}$ veto, and show that the cross section with the $\Tau_{Bj}$ veto is less sensitive than that with the $\Tau_{Cj}$ veto for the same central $p_{Tj}$ veto. Reduced sensitivity is of course advantageous, given that our current theoretical description of these effects is based on models rather than first principles theory.

This ordering of sensitivities to UE and hadronisation effects actually makes intuitive sense. The rapidity-dependent vetoes $\Tau_{B/Cj}$ impose a similar veto as $p_{Tj}$ at $y_j=Y$, but as one moves away from the Higgs rapidity and $|y_j-Y|$ increases, the veto is lifted and one moves towards simply measuring the inclusive Higgs cross section in these forward regions. The inclusive Higgs cross section is, of course, much less affected by hadronisation and UE than the cross section with a restriction on $p_{Tj}$, and this leads to the cross sections with a $\Tau_{B/Cj}$ veto being less sensitive to UE and hadronisation than that with a $p_{Tj}$ veto. With the same `central' $p_{Tj}$ veto at $y_j=Y$, the veto is lifted more quickly as one goes forward in rapidity for $\Tau_{Bj}$ than $\Tau_{Cj}$ (see Fig.~1 of Ref.~\cite{Gangal:2014qda}), and this leads to the cross section with a $\Tau_{Bj}$ veto being less sensitive to UE and hadronisation than the cross section with a $\Tau_{Cj}$ veto.

\section{Conclusions}
\label{sec:V}

In this paper, we obtained NNLL$'+$NNLO predictions for the 0-jet gluon-fusion Higgs cross section, $\sigma_0(\Tau^\cut)$, where the jet veto is imposed by requiring that no identified jet has a value of $\Tau_{B/Cj}$ greater than $\Tau^\cut$. The observables $\Tau_{Bj}$ and $\Tau_{Cj}$, defined in Eq.~\eqref{eq:TauBCdefs}, correspond to a rapidity-dependent jet veto: they impose the tightest constraint on jet transverse mass at `central' rapidities close to the Higgs rapidity, with the veto gradually loosening as one goes to forward rapidities away from the Higgs. The perturbative uncertainty in these predictions has been estimated through combined scale variations of the different resummation and fixed-order scales involved. We compared the NNLL$'+$NNLO predictions to lower-order NLL and NLL$'+$NLO ones, observing in general that the perturbative uncertainties significantly reduce as the perturbative order is increased, and that the predictions of higher orders fall within the uncertainty bands of lower orders, indicating good perturbative convergence. Explicit results have been provided for jet radius $R=0.5$, but results for other jet radii can be provided on request to the authors.

In the `resummation region' $\Tau^\cut \ll m_H$, the soft and beam resummation scales $\mu_S$ and $\mu_B$ in our predictions should be chosen to be $r_s\Tau^\cut$ and $\sqrt{r_s m_H \Tau^\cut}$ respectively, where $r_s$ is some number of order 1. We investigated the use of both $r_s=1$ and $r_s=2$. Whilst for $\Tau_{Bj}$ the $r_s=1$ and $r_s=2$ predictions look rather similar, for $\Tau_{Cj}$ use of $r_s=2$ notably improves the perturbative convergence and avoids an unphysical behaviour in the jet-vetoed cross section. For these reasons, we advocate the use of our $r_s=2$ results for $\Tau_{Cj}$. Taking $r_s=2$ for $\Tau_{Cj}$, we find, setting $\Tau^\cut$ to the benchmark value of $20$ GeV and $R$ to $0.5$, that the perturbative uncertainty on the vetoed cross section reduces from 15.7\% at NLL$'+$NLO to 9.77\% at NNLL$'+$NNLO, with the central value increasing from 33.4 pb to 37.3 pb. Repeating the exercise with $\Tau_{Bj}$ and taking $r_s=1$, we find a reduction in uncertainty from 21.2\% to 10.9\%, with the central value increasing from 32.9 pb to 37.0 pb.

Using an NLO+PS set-up, we compared the effect of underlying event (UE) and hadronisation on $0$-jet ggH cross sections where the jet veto constraint was implemented via $\Tau_{Bj}$, $\Tau_{Cj}$ or the conventional jet veto observable $p_{Tj}$. Adjusting all three vetoes such that they imposed the same jet veto at central rapidities (at small $R$), we found that the cross-section with the $\Tau_{Bj}$  veto was minimally sensitive to both UE and hadronisation, followed by that with $\Tau_{Cj}$, and then finally the cross section with the $p_{Tj}$ veto was the most sensitive. The fact that the cross sections with rapidity-dependent jet vetoes have a reduced sensitivity to these theoretically less-well-understood effects is one advantage of using such vetoes.

The use of $\Tau_{B/Cj}$ rather than $p_{Tj}$ to classify and veto jets has practical advantages, and also provides complementary information on the properties of additional jet production in a given hard process. We look forward to comparing our predictions for $0$-jet ggH cross sections with a $\Tau_{B/Cj}$ veto against data from the LHC experiments.

\begin{acknowledgments}

We would like to thank Rikkert Frederix, Valentin Hirschi, Maximilian Stahlhofen and Paolo Torrielli for useful discussions. SG acknowledges the CERN theory group for hospitality, where a part of this work was completed. FT acknowledges support from the Deutsche Forschungsgemeinschaft (DFG) under Germany's Excellence Strategy -- EXC 2121 ``Quantum Universe'' 
-- 390833306.

\end{acknowledgments}

\bibliographystyle{JHEP}
\bibliography{TaujetNNLL}
\end{document}